\documentclass[10pt,journal,twocolumn]{IEEEtran}
\usepackage{setspace}
\IEEEoverridecommandlockouts
\usepackage{cite}
\usepackage{amsmath,amssymb,amsfonts,booktabs}
\usepackage{algorithmic}
\usepackage{graphicx}
\usepackage{textcomp}
\usepackage{xcolor}
\usepackage{enumerate}
\usepackage{epstopdf}
\usepackage{subcaption}
\usepackage{amsthm}
\newtheorem{theorem}{Theorem}

\newtheorem{proposition}{Proposition}

\usepackage{multirow} % Required for multirows

\definecolor{mygray}{RGB}{240,240,240}
\usepackage{colortbl}

\usepackage{algorithm}
\usepackage{algorithmic}
\allowdisplaybreaks[4]

\begin{document}

\title{Energy-Efficient Multi-UAV-Enabled MEC Systems over Space-Air-Ground Integrated Networks}

\author{
        Wenchao Liu,
        Xuhui Zhang,
        Chunjie Wang,
        Jinke Ren,
        Zheng Xing,
        Bo Yang,
        Shuqiang Wang,
        and Yanyan Shen

\thanks{
Wenchao Liu is with the School of Automation and Intelligent Manufacturing, Southern University of Science and Technology, Guangdong 518055, China (e-mail: wc.liu@foxmail.com).
}

\thanks{
Xuhui Zhang, Jinke Ren, and Zheng Xing are with the Shenzhen Future Network of Intelligence Institute, the School of Science and Engineering, and the Guangdong Provincial Key Laboratory of Future Networks of Intelligence, The Chinese University of Hong Kong, Shenzhen, Guangdong 518172, China (e-mail: xu.hui.zhang@foxmail.com; jinkeren@cuhk.edu.cn; zhengxing@link.cuhk.edu.cn).
}

\thanks{
Chunjie Wang, Shuqiang Wang, and Yanyan Shen are with the Shenzhen Institute of Advanced Technology, Chinese Academy of Sciences, Guangdong 518055, China, and also with Shenzhen University of Advanced Technology, Guangdong 518055, China (e-mail: cj.wang@siat.ac.cn; sq.wang@siat.ac.cn; yy.shen@siat.ac.cn).
}

\thanks{B. Yang is with the Department of Automation and the Key Laboratory of System Control and Information Processing, Ministry of Education, Shanghai Jiao Tong University, Shanghai 200240, China (e-mail: bo.yang@sjtu.edu.cn).
}

}
\maketitle

\begin{abstract}
With the development of artificial intelligence integrated next-generation communication networks, mobile users (MUs) are increasingly demanding the efficient processing of computation-intensive and latency-sensitive tasks. However, existing mobile computing networks struggle to support the rapidly growing computational needs of the MUs. Fortunately, space-air-ground integrated network (SAGIN) supported mobile edge computing (MEC) is regarded as an effective solution, offering the MUs multi-tier and efficient computing services.
In this paper, we consider an SAGIN supported MEC system, where a low Earth orbit satellite and multiple unmanned aerial vehicles (UAVs) are dispatched to provide computing services for MUs. An energy efficiency maximization problem is formulated, with the joint optimization of the MU-UAV association, the UAV trajectory, the task offloading decision, the computing
frequency, and the transmission power control.
Since the problem is non-convex, we decompose it into four subproblems, and propose an alternating optimization based algorithm to solve it.
Simulation results confirm that the proposed algorithm outperforms the benchmarks.
\end{abstract}

\section{INTRODUCTION}
\IEEEPARstart{T}{he} advancements in mobile communication networks have led to an exponential increase in the demand for ubiquitous access from mobile users (MUs) \cite{Liu2018Space-Air-Ground}. Despite the significant enhancement of terrestrial network convenience afforded by the fifth-generation (5G) technology, connectivity issues, such as weak wireless access capability and unsatisfied computing services, still exist in areas with sparse MUs,
due to the limitations in coverage and economic costs \cite{Zhang2023Multiagent}. To address these issues and achieve pervasive connectivity in remote regions, the integration of non-terrestrial networks has emerged as a promising strategy in the next-generation communication systems \cite{Wang2021Survey, 10980172} to provide ubiquitous services for the MUs.

Space–air–ground integrated network (SAGIN), as a typical non-terrestrial network, includes three primary components: the space-based network, which includes low Earth orbit (LEO) satellites; the air-based network, encompassing aircraft, unmanned aerial vehicles (UAVs), and high-altitude platforms; and the ground-based network, consisting of terrestrial base stations (BSs), MUs, and Internet of Things (IoT) devices \cite{Liu2020Task-Oriented}.
Satellite-enabled non-terrestrial networks can provide the MUs with extensive connectivity across both territory and time \cite{9508471}. However, due to the long distance between the satellites and the ground, it faces challenges such as high transmission latency, low transmission rates, and increased energy consumption for long-distance transmissions.
In contrast, terrestrial communication performs excellently in urban areas with the high density of MUs, providing high-speed and low-latency services \cite{7901477}. However, due to its limited coverage, it struggles to meet the communication needs of MUs located in remote regions \cite{Wang2024Hybrid}.
To tackle this issue, the UAVs can serve as aerial communication devices to provide communication and computing services. Due to their agile mobility, the UAVs can offer more flexible coverage and more stable communication quality for the MUs \cite{Wu2018Joint, 10972043}.
Through this multi-tier connectivity of the SAGIN, the MUs can benefit from the diverse services offered by different service providers including the LEO satellites and the UAVs, thereby the quality of services of all MUs can be improved \cite{Dazhi2022Terminal-Aware,Bakambekova2024On}.

Furthermore, the widespread utilization of those famous computation-intensive and latency-sensitive tasks and devices, including the multimedia applications, social networks, and emerging intelligent devices such as wearable devices and smart home controllers, has significantly impacted the MUs.
The increase in usage has heightened the demand for efficient communication connectivity and high-speed data transmission.
Additionally, it has highlighted the importance of enhancing the processing capabilities of the MUs to effectively manage the growing task loads.
As a result, optimizing task processing performance has become a critical issue deserving urgent attention \cite{Liu2023Energy-Efficient,Zhang2024A}. This requires efficient and reliable communication and computing resources. However, traditional cloud-centric computing models struggle to provide the MUs located at remote areas with high-quality user experiences. To address this issue, mobile edge computing (MEC) has been proposed as a solution that efficiently utilizes computational resources at the network edge \cite{Mao2017ASurvey}. The MEC brings computing and storage resources to the edge of the mobile network, placing them closer to the ground MUs. Through localized data processing and real-time analysis, the energy consumption and latency of task processing can be reduced, and the adaptability of the network is also enhanced \cite{Mach2017Mobile}.

However, effectively utilizing computing resources in the SAGIN supported MEC systems to enhance task processing efficiency remains a topic requiring further investigation to address various challenges. Firstly, the maximum processing latency of tasks must be considered to ensure that the data processing time
satisfies the latency requirements.
Secondly, the MUs and the aerial computing platforms such as the UAVs and high-altitude platforms often have limited energy capacity, necessitating the optimized energy usage to extend their battery lifetimes. Additionally, the dynamic nature of the network increase the complexity of resource allocation and task scheduling, requiring efficient algorithms to address these challenges.

To the best of our knowledge, there is a lack of study on the hierarchical computation offloading in SAGIN that considers the MU-UAV association and resource allocation for multiple UAVs with their controllable flying trajectories, while meeting the latency requirements of the MUs.
To fill this research gap, this article investigates an SAGIN-supported MEC system that jointly considers the MU-UAV association, the multiple UAVs trajectory design (TD), transmit power allocation, task offloading decisions, and computing frequency allocation.
The main contributions can be summarized as follows:
\begin{itemize}
    \item Firstly, we propose a novel SAGIN-supported MEC system with multiple UAVs, aiming at achieving extensive coverage and hotspot communication enhancement. Under the constraints of maximum latency of the MU tasks and maximum energy consumption of the UAVs, we aim to maximize the total energy efficiency of the system through jointly optimizing the MU-UAV association, the UAV trajectory, task offloading decision, transmit power control and computing frequency control.
    \item Secondly, due to the coupling between variables, the binary nature of the MU-UAV association, and the fractional structure in the objective function, the initial problem becomes a non-convex mixed-integer non-linear optimization problem. We first utilize quadratic transformation to eliminate the fractional structure, then apply an alternating optimization (AO)-based algorithm to decompose and solve the problem.
    \item Numerical results illustrate the impact of different parameters in the considered SAGIN-supported MEC system, such as the number of MUs, the data size of tasks required for processing, and the bandwidth among the MUs and the UAVs, on the achievable performance. The results also highlight the benefits of optimizing the UAV trajectories, optimizing the offloading allocations, and adjusting the number of UAVs. Additionally, the results display the UAV trajectories and the changes in task allocation over time.
\end{itemize}

The remainder of this article is organized as follows. In Section \ref{section2}, we discuss the existing related works in details. Section \ref{section3} introduces the system model, 
and provides the problem formulation. In Section \ref{section4}, we introduce the proposed AO-based algorithm and offers convergence and complexity analysis. Section \ref{section5} evaluates the performance of the proposed algorithm through numerical analysis. Finally, in
Section \ref{section6}, we summarize this article.

\section{RELATED WORKS}\label{section2}
Currently, some pioneering works began to explore the deployment of SAGIN in next-generation communication networks \cite{9953964, 9380358, 10458883, 9606690, 9963692, 9918062, 10233456, 10454605, 9964037, 11153428, 11153721}.
Benefited from the global coverage capabilities of the LEO satellites, the SAGIN can provide a new approach for achieving seamless coverage and enhancing the ubiquitous computing services.
Specifically, Chen et al. \cite{9953964} proposed an intelligent user association strategy in a novel SAGIN system, and the capacity and coverage probability are also analysed.
Wang et al. \cite{9380358} exploited the target positioning in a maritime SAGIN system, where the direction-of-arrival of the targets were estimated by the UAVs.
Mohamed et al. \cite{10458883} studied the joint UAV trajectory planning and LEO satellite selection, and the user data rate was maximized.
Tang et al. \cite{9606690} proposed a learning based network traffic control for SAGIN by considering the high mobility of users as well as frequent changing network traffic and link state.
Besides, the joint deployment and optimization of the SAGIN and next-generation communication technologies, including the backscatter communications \cite{9963692}, the blockchain-enabled systems \cite{9918062, 10233456}, the IoT data collections \cite{10454605}, and the reconfigurable intelligent surface-assisted systems \cite{9964037}, have been studied.
{
Xu et al. \cite{11153428} investigated the energy-efficient optimization problem in an SAGIN system for vehicular communications.
Han et al. \cite{11153721} studied the data freshness optimization problem for the wireless power tranfer-assisted data collection system over the SAGIN.
}
However, the increasing demand for the computing tasks by the emerging applications from the MUs still cannot be satisfied.

Fortunately, the MEC system, as a new distributed computing paradigm based on wireless communication networks, has attracted widespread attentions.
Several pioneer works \cite{8387798, 8664595, 8488502, 8960510, 9417469, 10373153, 10606316, 10417719} started to investigate the computing potential of the implementation of the MEC systems towards next-generation wireless communication networks.
Ren et al. investigated the resource allocation \cite{8387798}, and the collaboration of cloud computing and edge computing \cite{8664595} to minimize the latency of all mobile devices.
Cao et al. \cite{8488502} proposed a new cooperation method in both computation and communication resources for MEC systems to improve the energy efficiency.
Wang et al. \cite{8960510} studied a wireless powered MEC system, where the total transmission energy consumption was minimized considering both energy and task causality constraints.
{
Li et al. \cite{8876867} discussed the online incentive mechanism design integrating computation and communication resource allocation for collaborative task offloading in an MEC system.
Yi et al. \cite{8606230} investigated the multi-user task offloading and scheduling considering tradeoffs between the local computing and the edge computing for a multi-access MEC system.
}
Zhang et al. \cite{9417469} optimized the sum of utilities, which jointly considered the energy efficiency, time latency, and price of offloading computations in an intelligent reflecting surface aided MEC system.
Liang et al. \cite{10373153} studied the joint task offloading, communication, and computation resource allocation in a multi-user MEC system.
Meanwhile, the UAV-enabled MEC systems were studied in \cite{10606316, 10417719}, where the systems are benefited by the mobility of the UAV through optimizing the UAV trajectories and resource allocations.
Nevertheless, these works mainly focused on the terrestrial networks. The MUs in remote areas are still unable to utilize cloud and edge computing resources to handle computational tasks due to the lack of effective network access.

Introducing the SAGIN into the MEC systems has become a key emerging approach to address the aforementioned issues.
Notably, the LEO satellites provide extensive global coverage, while the UAVs offer flexible deployment and dynamic response capabilities.
This combination not only compensates for the limitations of single networks but also delivers more stable and efficient computing and communication services across various application scenarios, significantly enhancing task processing efficiency.
There has been several works dedicated to developing strategies for efficient task processing and computing resource management in SAGIN-supported MEC systems \cite{10440193, Hu2023Joint, 10342725, 10579794, 10879508, 10891825, 10947633, 11134095, 11145097, 11192086, 11124243}.
Specifically, Huang et al. \cite{10440193} studied the minimization of total energy consumption for task processing in a SAGIN-supported MEC system.
Hu et al. \cite{Hu2023Joint} focused on the maximization of total energy efficiency by jointly optimizing the trajectories of two UAVs with different operations, computing resource allocation, and bandwidth allocation.
Nguyen et al. \cite{10342725} studied the computation offloading problem in hybrid edge-cloud based SAGIN, where the total system energy consumption was minimized.
Du et al. \cite{10579794} presented an MEC and blockchain enabled SAGIN architecture, while minimized the energy consumption through the task segmentation, the UAVs and satellite’s bandwidth allocation among their served IoT devices.
{Zhu et al. \cite{10879508} jointly optimized system energy consumption and delay costs in an SAGIN framework by designing the resource allocation, the task offloading, and the channel allocation.
Li et al. \cite{10891825} studied the offloading optimization and bandwidth allocation in a joint UAV-LEO satellite offloading SAGIN.
Xie et al. \cite{10947633} aimed to optimize the delay and fair utility of the SAGIN and improve the utilization of the UAVs and the satellites.
}
{
Zhang et al. \cite{11134095} studied the UAV deployment and the computation offloading optimization for the energy-aware SAGIN.
Hsu et al. \cite{11145097} proposed a service caching
and task offloading scheme for the MEC service optimization with SAGIN.
Liu et al. \cite{11192086} investigated the latency minimization problem for the IoT system over SAGINs.
From a task offloading ratio perspective, Huynh et al. \cite{11124243} optimized the bandwidth allocations and the task offloading portions to improve the quality of the MEC services.
}

However, existing works critically neglect four fundamental challenges in multi-UAV-enabled MEC systems over SAGIN: (1) the absence of hybrid access mechanism optimization for enhanced spectral efficiency; (2) insufficient joint resource coordination in hierarchical communication and computation optimization (HCCO), scheduling (e.g., LEO satellite and UAVs), and the device-side local processing of MUs, to fully utilize the computation capability; (3) the absence of energy-efficiency optimization (EEO) with TD across multiple UAVs for flight energy conservation; (4) unexplored flexible channel-aware dynamic association strategies (DASs) between UAVs and MUs to enable precision offloading decisions.
These unresolved challenges constrain energy efficiency improvement for the whole SAGIN-supported MEC system.
{
To clearly illustrate our contributions, Table \ref{trw} summarizes the key differences between our work and prior works related to the SAGIN-supported MEC systems, particularly in terms of the system design and the optimization objective.
}

\begin{table}[!htbp] \footnotesize
	\centering
	\caption{{Comparison of Existing Works}}
	\label{trw}
	\begin{tabular}
    % {>{\centering}m{1cm}
    % >{\centering\arraybackslash}c
    % >{\centering\arraybackslash}p{1cm}
    % >{\centering\arraybackslash}p{1cm}
    % >{\centering\arraybackslash}c
    % >{\centering\arraybackslash}c}
    {
    m{1.9cm}<{\centering}
    m{0.815cm}<{\centering}
    m{0.815cm}<{\centering}
    m{0.45cm}<{\centering}
    m{0.525cm}<{\centering}
    m{0.525cm}<{\centering}
    m{0.63cm}<{\centering}
    }
  \toprule	
  Works & Hybrid Access & Multiple UAVs & TD 
 & DAS & EEO & HCCO \\	
  \midrule  %
  \cite{10440193, 10342725} & $\times$ & $\surd$ & $\surd$ & $\surd$ & $\times$ & $\times$ \\
  \cite{Hu2023Joint} & $\surd$ & $\surd$ & $\surd$ & $\surd$ & $\surd$ & $\times$ \\
    \cite{10579794, 10879508, 10947633} & $\times$ & $\surd$ & $\times$ & $\times$ & $\times$ & $\times$ \\ 
  \cite{10891825} & $\times$ & $\surd$ & $\surd$ & $\times$ & $\surd$ & $\times$ \\
    \cite{11134095, 11145097, 11124243} & $\times$ & $\surd$ & $\times$ & $\surd$ & $\times$ & $\times$ \\
  \cite{11192086} & $\times$ & $\surd$ & $\surd$ & $\times$ & $\times$ & $\times$ \\
  \cellcolor{mygray}{\textbf{Ours}} & \cellcolor{mygray}{$\boldsymbol{\surd}$} & \cellcolor{mygray}{$\boldsymbol{\surd}$} & \cellcolor{mygray}{$\boldsymbol{\surd}$} & \cellcolor{mygray}{$\boldsymbol{\surd}$} & \cellcolor{mygray}{$\boldsymbol{\surd}$} & \cellcolor{mygray}{$\boldsymbol{\surd}$}  \\
		\bottomrule\\
        % \multicolumn{7}{c}{EEO: energy efficiency optimization.}
	\end{tabular}
\end{table}

\section{SYSTEM MODEL AND PROBLEM FORMULATION}\label{section3}
\begin{figure}[!htbp]
	\centering
	\includegraphics[width=1\linewidth]{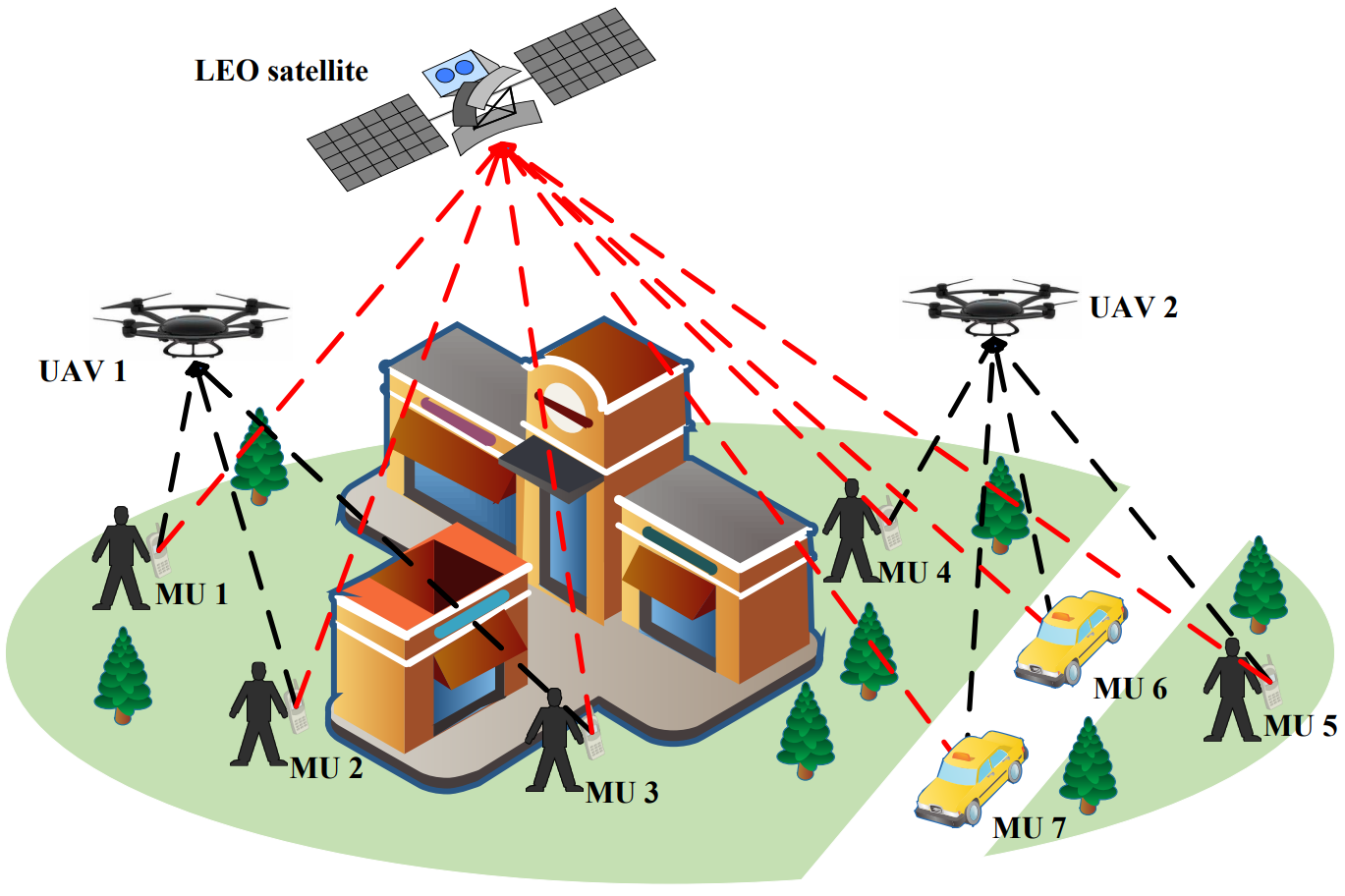}
	\caption{System model of the space-air-ground integrated MEC.}
	\label{system model}
\end{figure}

As depicted in Fig. \ref{system model}, we consider an SAGIN-supported MEC system, which consists of $M$ MUs, denoted as $\mathcal{M} = \{1,2,\cdots, M \}$, $K$ UAVs, denoted as $\mathcal{K} = \{1,2,\cdots, K \}$, and an LEO satellite.
The LEO satellite and each UAV are equipped with an onboard computing server, which can provide computing services to the MUs.
Meanwhile, each MU has a computation task, which can be separated into several independent and fine-grained subtasks \cite{Zeng2023Joint}.
Due to the computing latency requirements of the tasks and the limited computational capabilities of the MUs, a portion of these tasks need to be offloaded to the UAVs and the LEO satellite for remote computing.\footnote{
A promising future direction lies in integrating digital twin (DT) and AI technologies with the MEC systems over the SAGINs \cite{10238695}. DTs can create dynamic virtual replicas of the SAGINs, enabling predictive resource management and intelligent orchestration for heterogeneous tasks \cite{10234396, 10540318}, particularly in emerging applications such as personalized healthcare and human-virtual interaction systems.
}

We consider a time period $\mathcal{T}$ with $N$ time slots, denoted as $\mathcal{N}= \{1,2,\cdots,N\}$, with each time slot having equal duration $\tau$. At the beginning of each time slot, a task will be generated at each MU. The hybrid time division and frequency division multiple access is considered \cite{Nguyen2022Joint}.
Since the duration $\tau$ is very small, the position of the UAV during each time slot is fixed, while varies across different time slots.
It is assumed that the UAVs fly at a fixed altitude $H$, which enables them to fly smoothly and avoid obstacles and buildings.
Hence, the position of the UAV $k$ can be denoted as  ${\mathcal{Q}}_{k}[n] = (\boldsymbol{q}_{k}[n], H)^{\mathsf{}}$, where $\boldsymbol{q}_k[n] = ( x_k[n] , y_k[n] )^{\mathsf{}}$ is the horizontal coordinates.
Moreover, the location of the MU $m$ is denoted as ${\mathcal{S}}_{m}[n] = (\boldsymbol{s}_{m}[n], 0)^{\mathsf{}}$, where $\boldsymbol{s}_m[n] = (x_m[n] , y_m[n] )^{\mathsf{}}$.

\subsection{UAV KINETIC MODEL}
The initial and final positions of each UAV are defined as $\boldsymbol{q}_k[1] = \boldsymbol{q}_{k,\mathsf{I}},
 \boldsymbol{q}_k[N] = \boldsymbol{q}_{k,\mathsf{F}}, \forall k$, respectively.
To ensure the flight stability of all UAVs, the constraint that avoids the collisions among the UAVs, and the constraint for the UAV speed must be satisfied, which are given by
\begin{equation}
	d_{\min}^{2} \le \| \boldsymbol{q}_k[n] -\boldsymbol{q}_i[n] \|^{2},\ 
\forall k, i\in \mathcal{K},i \neq k, \forall n \in \mathcal{N},
\end{equation}
\begin{equation}
\| \boldsymbol{v}_{k}[n] \| = \frac{ \| \boldsymbol{q}_k[n] -\boldsymbol{q}_k[n-1] \|}{\tau}  \le V_{\max},\ \forall k, \forall n \in \mathcal{N}\backslash\{1\},
\end{equation}
where $d_{\min}$ denotes the minimum safe distance between any two UAVs, and $V_{\max}$ denotes the maximum speed of the UAVs.
Furthermore, the energy consumption for the flight of the UAV $k$ can be expressed as
\begin{equation}
E_{k}^{\rm{prop}}[n] = \tau\left(\zeta_1\|\boldsymbol{v}_{k}[n]\|^3+\frac{\zeta_2}{\|\boldsymbol{v}_{k}[n]\|}\right),
\end{equation}
where $\zeta_1$ and $\zeta_2$ are fixed parameters related to the UAV’s weight, wing area, air density, etc \cite{Zeng2017Energy-Efficient}.

\subsection{COMMUNICATION MODEL}
We assume that each MU is equipped with two communication interfaces: one for communicating with the LEO satellite and another for communicating with the associated UAV. The two interfaces operate on distinct spectral bands, ensuring that there is no interference between the communications with the UAVs and the communications with the LEO satellite \cite{Zhang2014Dynamic,Zhou2021Deep}.

Due to the flexibility of the UAVs, they can fly close to MUs to establish line-of-sight (LoS) links \cite{Diao2020Fairness-Aware, Zeng2017Energy-Efficient, Nguyen2022Joint, Wu2018Joint, 10606316}. The channel gain between the MU $m$ and UAV $k$ can be expressed as
\begin{equation}
h_{m,k}[n] = \frac{ \beta_{0}}{d_{m,k}^{2}[n]} = \frac{\beta_{0}}{H^{2} + \| 
\boldsymbol{q}_k[n] - \boldsymbol{s}_m[n] \|^{2} }, 
\end{equation}
where $\beta_{0} $ represents the channel gain at the reference distance $d_0 = 1$m.
Thus, the transmission rate from MU $m$ to UAV $k$ can be expressed as
\begin{equation}
R_{m,k} [n] = B_{m,k} \log_2 \left( 1 + \frac{p_{m,k}[n] h_{m,k} [n]}{\sigma_{{\rm{UAV}}}^2} \right),
\end{equation}
where $B_{m,k}$ denotes the bandwidth allocated for data transmission between MU $m$ and UAV $k$, $p_{m,k}[n]$ is the transmit power of MU $m$ for communicating with UAV $k$, and $\sigma_{{\rm{UAV}}}^2$ is the noise power.

Besides, a portion of the task of MUs can be offloaded to the LEO satellite for computing.
Accordingly, the channel conditions between the MUs and the LEO satellite are primarily influenced by the distance and meteorological environment. We assume that the meteorological environment remains constant during the task processing period, hence the channel gain between each MU and the LEO satellite is predominantly determined by the satellite's altitude\cite{Liu2023Energy-Efficient,Zhang2023Learning,Zhou2021Deep}.
Thus, the transmission rate between the MU $m$ and the LEO satellite can be expressed as
\begin{equation}
R_{m,{\rm{LEO}}}[n] = B_{{\rm{LEO}}} \log_2 \left( 1 + \frac{p_{m,{\rm{LEO}}}[n] \left|g_{m, {\rm{LEO}}}\right|^{2}}{\sigma_{{\rm{LEO}}}^2} \right),
\end{equation}	
where $B_{{\rm{LEO}}} $ denotes the bandwidth allocated for data transmission between each MU and the LEO satellite, $p_{m,{\rm{LEO}}}$ denotes the transmit power of the MU $m$ for communicating with the LEO satellite, $g_{m, {\rm{LEO}}}$ is the channel gain between the MU $m$ and the LEO satellite, and $\sigma_{{\rm{LEO}}}^2$ is the noise power.

\subsection{COMPUTING MODEL}
At the beginning of each time slot, each MU arrives a task, denoted as $\{ D_{m}[n], \phi_{m}[n] \}$, where $D_{m}[n]$ represents the data size of the task, and $\phi_{m}[n]$ represents the
computing cycles required to process one bit data.
Let $ \{ \omega_{m}^{\rm{L}}[n], \omega_{m}^{\rm{U}}[n],\omega_{m}^{\rm{S}}[n]\}$ denote the set of task offloading decision of the MU $m$, where $ \omega_{m}^{\rm{L}}[n], \omega_{m}^{\rm{U}}[n],\omega_{m}^{\rm{S}}[n] \in [0, 1] $ represent the proportions of tasks processed locally, offloaded to the associated UAV, and offloaded to the LEO satellite, respectively. The task offloading decision satisfies $\omega_{m}^{\rm{L}}[n] + \omega_{m}^{\rm{U}}[n] + \omega_{m}^{\rm{S}}[n] = 1, \forall m,n$.

\subsubsection{Local Processing}
Let $f^{\rm{L}}_{m}[n]$ denote the number of computing cycles per second of the MU $m$ at time slot $n$. Then, the latency for processing $\omega_{m}^{\rm{L}}[n]D_{m}[n]$-bit task locally at the MU $m$ is 
\begin{equation}
L_{m}^{\rm{L}}[n] = \frac{ \omega_m^{\rm{L}}[n] D_{m}[n] \phi_{m}[n] }{f^{\rm{L}}_{m}[n]}.
\end{equation}
Accordingly, the energy consumption for local processing is 
\begin{equation}
E_{m}^{\rm{L}}[n] = \kappa_{\rm{L}} (f^{\rm{L}}_{m}[n])^{2} \omega_m^{\rm{L}}[n] D_{m}[n] \phi_{m}[n],
\end{equation}
where $\kappa_{\rm{L}}$ denotes the energy efficiency factor for the local computing of all MUs.

\subsubsection{UAV Processing}
When the MU $m$ offloads a partial task $\omega_{m}^{\rm{U}}[n]D_{m}[n]$ to the associated UAV, the latency
comprises three parts:
the time for task data uplink transmission from the MU to the associated UAV, the time for the task computation at the associated UAV, and the time for task computation result downlink transmission from the UAV to the MU.
Since the data volume of the results is much smaller than that of the offloaded task, the download latency can be neglected \cite{Liu2023Energy-Efficient}.

Each MU is associated with a specified UAV per time slot, and a portion of their tasks can be offloaded to the associated UAV for computing. 
We introduce a binary indicator $\alpha_{m,k}[n]$ to denote the association, where $\alpha_{m,k}[n] = 1$ indicates that the MU $m$ is associated with the UAV $k$ at time slot $n$, otherwise $\alpha_{m,k}[n] = 0$.
Let $f^{\rm{U}}_{m,k}[n]$ denote the number of computing cycles per second that UAV $k$ allocates to MU $m$. The latency for completing the $\omega_{m}^{\rm{U}}[n]D_{m}[n]$-bit task at the UAV $k$ consists of two parts: the task offloading and the task processing. Therefore, the latency can be expressed as
\begin{equation} \small
\begin{split}
    L_{m}^{\rm{U}}[n] = \sum_{k=1}^{K}\alpha_{m,k}[n] \Bigg( \frac{\omega_{m}^{\rm{U}}[n]D_{m}[n]}{R_{m,k} [n]} + 
\frac{ \omega_m^{\rm{U}}[n] D_{m}[n] \phi_{m}[n] }{f^{\rm{U}}_{m,k}[n]} \Bigg).
\end{split}
\end{equation}
Hence, the energy consumption of MU $m$ for offloading task to the associated UAV and the energy consumption for processing the task by UAV $k$ are given respectively by
\begin{equation}
E_{m}^{\rm{U},\rm{tran}}[n] =  \sum_{k=1}^{K}\alpha_{m,k}[n] \frac{p_{m,k}[n]\omega_{m}^{\rm{U}}[n]D_{m}[n]}{R_{m,k} [n]} ,
\end{equation}
\begin{equation}
E_{m,k}^{\rm{U},\rm{com}}[n] = \alpha_{m,k}[n] \kappa_{\rm{U}} (f^{\rm{U}}_{m,k}[n])^{2} \omega_m^{\rm{U}}[n] D_{m}[n] \phi_{m}[n],
\end{equation}
where $\kappa_{\rm{U}}$ denotes the energy efficiency factor for task processing of the UAVs.

\subsubsection{LEO Satellite Processing}
Similar to the tasks processed by the UAVs, the latency of the tasks processed by the LEO satellite also consists of two parts: the latency for offloading tasks from the MU to the LEO satellite and the latency for task computation at the LEO satellite.
Let $f^{\rm{S}}_{m}[n]$ denote the number of computing cycles per second that the LEO satellite allocates to MU $m$. The latency for processing $\omega_{m}^{\rm{S}}[n]D_{m}[n]$-bit task from MU $m$ at the LEO satellite can be expressed as
\begin{equation}
L_{m}^{\rm{S}}[n] =   \frac{\omega_{m}^{\rm{S}}[n]D_{m}[n]}{R_{m,\rm{LEO}} [n]} +
\frac{ \omega_m^{\rm{S}}[n] D_{m}[n] \phi_{m}[n] }{f^{\rm{S}}_{m}[n]}. 
\end{equation}
Accordingly, the energy consumption for task offloading from MU $m$ to the LEO satellite and the energy consumption for processing the task are given by
\begin{equation}
E_{m}^{\rm{S},\rm{tran}}[n] =  \frac{p_{m,\rm{LEO}}[n]\omega_{m}^{\rm{S}}[n]D_{m}[n]}{R_{m,\rm{LEO}} [n]} ,
\end{equation}
\begin{equation}
E^{\rm{S},\rm{com}}[n] = \kappa_{\rm{S}} (f^{\rm{S}}_{m}[n])^{2} \omega_m^{\rm{S}}[n] D_{m}[n] \phi_{m}[n],
\end{equation}
where $\kappa_{\rm{S}}$ denotes the energy efficiency factor for task computing of the LEO satellite.

To this end, the total latency and total energy consumption
of MU $m$ are given by
\begin{equation}
L_{m}^{{\rm{sum}}}[n] = \max \left\{ L_{m}^{\rm{L}}[n] , L_{m}^{\rm{U}}[n] , L_{m}^{\rm{S}}[n]  \right \},
\end{equation}
\begin{equation}
\begin{split}
E_{m}^{{\rm{sum}}}[n] =&\ E_{m}^{\rm{L}}[n] + E_{m}^{\rm{U},\rm{tran}}[n] + \sum_{k \in \mathcal{K}}E_{m,k}^{\rm{U},\rm{com}}[n] \\
&+ E_{m}^{\rm{S},\rm{tran}}[n] + E^{\rm{S},\rm{com}}[n].   
\end{split}
\end{equation}

\subsection{PROBLEM FORMULATION}
In this work, we aim to maximize the total energy efficiency of the system by jointly optimizing the UAV trajectory $\boldsymbol{Q} = \{\boldsymbol{q}_k[n], \forall k,n \}$, the transmit power of all MUs $\boldsymbol{P} = \{p_{m,k}[n],p_{m,{\rm{LEO}}}[n], \forall m,k,n \}$, the task offloading decision of all MUs $\boldsymbol{\Omega} = \{ \omega_{m}^{\rm{L}}[n], \omega_{m}^{\rm{U}}[n],\omega_{m}^{\rm{S}}[n], \forall m,n \}$, the association among the MUs and the UAVs $\boldsymbol{\alpha}=\{ \alpha_{m,k}[n], \forall m,k,n\}$, and the computing frequency allocation of the MUs, the UAVs and the LEO satellite $\boldsymbol{F} = \{f^{\rm{L}}_{m}[n], f^{\rm{U}}_{m,k}[n], f^{\rm{S}}_{m}[n], \forall m,k,n \}$.
The problem can be formulated as
\begin{subequations}
\begin{align}
\textbf{P1:} \ & \max\limits_{\{\boldsymbol{Q}, \boldsymbol{P}, \boldsymbol{\Omega}, \boldsymbol{F}, \boldsymbol{\alpha} \}} \sum_{m=1}^{M}\sum_{n=1}^{N} \frac{D_{m}[n]}{E_{m}^{{\rm{sum}}}[n]} \\
{\rm{s.t.}}\ &\boldsymbol{q}_k[1] = \boldsymbol{q}_{k,\mathsf{I}},
 \boldsymbol{q}_k[N] = \boldsymbol{q}_{k,\mathsf{F}},\forall k,\label{P11}\\
&d_{\min}^{2} \le \| \boldsymbol{q}_k[n] -\boldsymbol{q}_i[n] \|^{2}, 
\forall k, i\in \mathcal{K},i \neq k, \forall n,\label{P12}\\
& \| \boldsymbol{v}_{k}[n] \| \le V_{\max} ,\forall k, \forall n \in \mathcal{N}\backslash\{1\},\label{P13}\\
&\alpha_{m,k}[n] \in \left\{0,1\right\},\forall m,k,n, \label{P14}\\
&\sum_{k \in \mathcal{K}} \alpha_{m,k}[n] = 1, \forall m,n, \label{P15}\\
& \{ \omega_{m}^{\rm{L}}[n], \omega_{m}^{\rm{U}}[n],\omega_{m}^{\rm{S}}[n] \} \in [0,1],\forall m,n,\label{P16}\\
&\omega_{m}^{\rm{L}}[n] + \omega_{m}^{\rm{U}}[n] + \omega_{m}^{\rm{S}}[n] = 1, \forall m,n,\label{P17}\\
& 0 \leq f^{\rm{L}}_{m}[n] \leq F^{\rm{L}}_{\rm{max}}, \forall m,n, \label{P18}\\
& 0 \leq f^{\rm{U}}_{m,k}[n], \sum_{m \in \mathcal{M}} f^{\rm{U}}_{m,k}[n] \leq F^{\rm{U}}_{\rm{max}}, \forall m,k,n, \label{P19}\\
& 0 \leq f^{\rm{S}}_{m}[n], \sum_{m \in \mathcal{M}}f^{\rm{S}}_{m}[n] \leq F^{\rm{S}}_{\rm{max}}, \forall m,n, \label{P110}\\
& 0 \leq p_{m,k}[n] \leq P_{\rm{max}}^{\rm{U}}, \forall m,k,n, \label{P111}\\
& 0 \leq p_{m,\rm{LEO}}[n] \leq P_{\rm{max}}^{\rm{S}}, \forall m,n, \label{P112}\\
& L_{m}^{{\rm{sum}}}[n] \le \tau,\forall n,m,\label{P113}\\
& \sum_{n \in \mathcal{N}} \left( E_{k}^{\rm{prop}}[n] + \sum_{m \in \mathcal{M}}E_{m,k}^{\rm{U},\rm{com}}[n] \right) \le E_{\rm{max}}^{{\rm{UAV}}}, \forall k,n,\label{P114}
\end{align}
\end{subequations}
where $F^{\rm{L}}_{\rm{max}}$, $F^{\rm{U}}_{\rm{max}}$, and $F^{\rm{S}}_{\rm{max}}$ are the maximum computing frequencies of the MUs, the UAVs, and the LEO satellite, respectively. $P_{\rm{max}}^{\rm{U}}$ and $P_{\rm{max}}^{\rm{S}}$ denote the maximum transmit power of the MUs for communicating with the associated UAV, and the LEO satellite, respectively. $E_{\rm{max}}^{{\rm{UAV}}}$ denotes the energy budget of the UAVs.
The constraints (\ref{P11})-(\ref{P13}) represent the mobility constraints of the UAVs. The constraints (\ref{P14})-(\ref{P15}) represent the association constraints among the MUs and the UAVs. The constraints (\ref{P16})-(\ref{P17}) represent the task offloading decision constraints. 
The constraints (\ref{P18})-(\ref{P110}) represent the constraints on the computing frequencies for local task processing, the computing frequencies for the UAV task processing, and the computing frequencies for the LEO satellite task processing. 
The constraints (\ref{P111})-(\ref{P112}) represent the transmit power allocation from the MUs to the UAVs, and the transmit power allocation from the MUs to the LEO satellite. 
The constraint (\ref{P113}) represents the latency constraint. Lastly, the constraint (\ref{P114}) represents the constraint on the energy consumption of the UAVs.

In problem \textbf{P1}, the objective function $\frac{D_{m}[n]}{E_{m}^{{\rm{sum}}}[n]}$ has a non-convex fractional expression involving multiple coupled variables and a binary variable. As a result, \textbf{P1} is a challenging mixed-integer nonlinear programming problem, which is difficult to be solved by conventional convex optimization methods.

\section{JOINT TRAJECTORY DESIGN, ASSOCIATION, AND RESOURCE ALLOCATION}\label{section4}
In this section, we employ a quadratic transformation method to transfer the fractional structure in the objective function to decouple the numerators and denominators. We then decompose problem \textbf{P1} into four subproblems: the MU-UAV association, UAV trajectory optimization, task offloading decision, and transmit power control and CPU frequency control.
The four subproblems are optimized in an alternate optimization manner.

\subsection{QUADRATIC TRANSFORMATION-BASED PROBLEM REFORMULATION}
Firstly, we employ a quadratic transformation method to reformulate the objective function.
\begin{theorem}\label{theorem1}
Given $M \times N$ pairs of non-negative functions $A_{m,n}(\bold{x}) : \mathbb{R}^d \rightarrow \mathbb{R}^+$ and positive functions $B_{m,n}(\bold{x}) : \mathbb{R}^d \rightarrow \mathbb{R}^{++}$, the sum-of-ratios problem
\begin{subequations}
	\begin{align}
    	\max\limits_{\{ \bold{x} \}} &\frac{1}{M}\frac{1}{N} \sum_{m=1}^{M}\sum_{n=1}^{N} \frac{A_{m,n}(\bold{x})}{B_{m,n}(\bold{x})} \\
		{\rm{s.t.}}\ & \bold{x} \in \mathcal{X},
	\end{align}
\end{subequations}
is equivalent to
\begin{subequations}
     \begin{align}
      	\max\limits_{\{\bold{x},\bold{y} \}} &\frac{1}{M}\frac{1}{N} \sum_{m=1}^{M}\sum_{n=1}^{N} \left( 2y_{m,n}\sqrt{A_{m,n}(\bold{x})}-y_{m,n}^{2}B_{m,n}(\bold{x}) \right) \\
      	{\rm{s.t.}}\ & \bold{x} \in \mathcal{X},
     \end{align}
\end{subequations}
where $\bold{x}$ is the variable, $\mathcal{X}$ denotes the feasible set, $\bold{y}$ represents the quadratic coefficients with size $M \times N$, and the $m$th row and the $n$th column of $\bold{y}$ is $y_{m,n}$.
	  
\begin{proof}
Please refer to \cite{Shen2018Fractional1}.
  % \cite{Shen2018Fractional1,Shen2018Fractional2}.
\end{proof}
\end{theorem}

According to Theorem \ref{theorem1}, problem \textbf{P1} can be transformed into the problem \textbf{P2}, which is shown as follows
\begin{subequations}
\begin{align}
\textbf{P2:} \ & \max\limits_{\{\boldsymbol{X},\bold{y} \}} \sum_{m=1}^{M}\sum_{n=1}^{N} \left( 2y_{m,n}\sqrt{D_{m}[n]}-y_{m,n}^{2}E_{m}^{{\rm{sum}}}[n] \right)\label{P2}\\
{\rm{s.t.}}\ &\text{(\ref{P11})}-\text{(\ref{P114})},
\end{align}
\end{subequations}
where $\boldsymbol{X} \triangleq \{\boldsymbol{Q}, \boldsymbol{P}, \boldsymbol{\Omega}, \boldsymbol{F}, \boldsymbol{\alpha} \}$, $\bold{y} = \{y_{m,n}, \forall m,n \}$ represents a matrix of size $M \times N$. 
In order to achieve the optimal energy efficiency,
$\bold{y}$ is updated alternately according to
\begin{equation}
y_{m,n} = \cfrac{\sqrt{D_{m}[n]}}{E_{m}^{{\rm{sum}}}[n]},\forall m,n,  \label{P21}
\end{equation}
and $\boldsymbol{X}$ in problem \textbf{P2} is solved thereafter. Since each element in the matrix $\bold{y}$ is non-decreasing after each iteration, it can be proved that the convergence is guaranteed.

The matrix $\bold{y}$ is initialized with a small constant for each element. After that, we embark on a search for a feasible solution set for $\boldsymbol{X}$.
Subsequently, we update the value of each element in $\bold{y}$ based on the obtained feasible $\boldsymbol{X}$. By repeating this process for a certain number of iterations, we can ultimately determine the optimal values of $\bold{y}^{*}$ and $\boldsymbol{X}^{*}$. 
\subsection{SUBPROBLEM 1: MU-UAV ASSOCIATION OPTIMIZATION}
By fixing the variables $\boldsymbol{Q}, \boldsymbol{P}, \boldsymbol{\Omega}, \boldsymbol{F}$, the subproblem to optimize the MU-UAV association is formulated as
\begin{subequations}
\begin{align}
\textbf{SP1:} \ & \min\limits_{\{\boldsymbol{\alpha} \}}  \sum_{m=1}^{M}\sum_{n=1}^{N}  \alpha_{m,k}[n] y_{m,n}^{2}  \mathcal{W}_{1}[n]  \label{P3}\\
{\rm{s.t.}}\ &\text{(\ref{P14})}-\text{(\ref{P15})}, \text{(\ref{P113})}-\text{(\ref{P114}),}
\end{align}
\end{subequations}
where
\begin{equation}
    \begin{split}
    \mathcal{W}_{1}[n] = &\sum_{k \in \mathcal{K}} \frac{p_{m,k}[n]\omega_{m}^{\rm{U}}[n]D_{m}[n]}{R_{m,k} [n]}\\
    &+ \sum_{k \in \mathcal{K}} \kappa_{\rm{U}} (f^{\rm{U}}_{m,k}[n])^{2} \omega_m^{\rm{U}}[n] D_{m}[n] \phi_{m}[n].
    \end{split}
\end{equation}
Problem \textbf{SP1} is a standard integer linear programming problem, which can be solved using conventional optimization methods, such as the branch-and-bound algorithm \cite{2013MunozJoint}.

\subsection{SUBPROBLEM 2: TASK OFFLOADING DECISION OPTIMIZATION}
By fixing the variables $\boldsymbol{Q}, \boldsymbol{P}, \boldsymbol{F}, \boldsymbol{\alpha} $, the subproblem to optimize the task offloading decision is formulated as
\begin{subequations}
\begin{align}
\textbf{SP2:} \ & \min\limits_{\{\boldsymbol{\Omega} \}}  \sum_{m=1}^{M}\sum_{n=1}^{N}  y_{m,n}^{2} E_{m}^{{\rm{sum}}}[n] \label{P4-A}\\
{\rm{s.t.}}\ &\text{(\ref{P16})}-\text{(\ref{P17})}, \text{(\ref{P113})}-\text{(\ref{P114}),}
\end{align}
\end{subequations}
where
\begin{equation}
\begin{split}
&E_{m}^{{\rm{sum}}}[n] = \omega_{m}^{\rm{L}}[n] \mathcal{W}_{2}[n] + \omega_{m}^{\rm{U}}[n] \mathcal{W}_{3}[n] + \omega_{m}^{\rm{S}}[n] \mathcal{W}_{4}[n],\\
&\mathcal{W}_{2}[n] =  \kappa_{\rm{L}} (f^{\rm{L}}_{m}[n])^{2}  D_{m}[n] \phi_{m}[n], \\
&\mathcal{W}_{3}[n] =  \sum_{k=1}^{K}\alpha_{m,k}[n] \frac{p_{m,k}[n]D_{m}[n]}{R_{m,k} [n]} \\
& \quad\quad\quad\quad + \sum_{k=1}^{K} \alpha_{m,k}[n] \kappa_{\rm{U}} (f^{\rm{U}}_{m,k}[n])^{2}  D_{m}[n] \phi_{m}[n],\\
&\mathcal{W}_{4}[n] = \frac{p_{m,\rm{LEO}}[n]D_{m}[n]}{R_{m,\rm{LEO}} [n]} +  \kappa_{\rm{S}} (f^{\rm{S}}_{m}[n])^{2} D_{m}[n] \phi_{m}[n].
\end{split}
\end{equation}
Problem \textbf{SP2} is a standard linear programming problem, which can be solved using the interior-point method \cite{2010LuoSemidefinite}.

\subsection{SUBPROBLEM 3: TRANSMIT POWER CONTROL AND COMPUTING FREQUENCY CONTROL OPTIMIZATION}
By fixing the variables $\boldsymbol{Q}, \boldsymbol{\Omega}, \boldsymbol{\alpha} $, the subproblem to optimize the transmit power control and computing frequency control is formulated as
\begin{subequations}
\begin{align}
\textbf{SP3:} \ & \min\limits_{\{\boldsymbol{P}, \boldsymbol{F} \}}  \sum_{m=1}^{M}\sum_{n=1}^{N}  y_{m,n}^{2} E_{m}^{{\rm{sum}}}[n] \label{P4-B}\\
{\rm{s.t.}}\ &\text{(\ref{P18})}-\text{(\ref{P114}).}
\end{align}
\end{subequations}
where $E_{m}^{{\rm{sum}}}[n] = \mathcal{W}_{5}[n] + \mathcal{W}_{6}[n]$, and given by
\begin{equation}
\begin{split}
\mathcal{W}_{5}[n] =&\ \kappa_{\rm{L}} (f^{\rm{L}}_{m}[n])^{2} \omega_m^{\rm{L}}[n] D_{m}[n] \phi_{m}[n]\\
&+ \kappa_{\rm{S}} (f^{\rm{S}}_{m}[n])^{2} \omega_m^{\rm{S}}[n] D_{m}[n] \phi_{m}[n]\\
&+ \sum_{k \in \mathcal{K}}\alpha_{m,k}[n] \kappa_{\rm{U}} (f^{\rm{U}}_{m,k}[n])^{2} \omega_m^{\rm{U}}[n] D_{m}[n] \phi_{m}[n],
\end{split}
\end{equation}
\begin{equation}
\begin{split}
\mathcal{W}_{6}[n] = &\ \frac{p_{m,\rm{LEO}}[n]\omega_{m}^{\rm{S}}[n]D_{m}[n]}{R_{m,\rm{LEO}} [n]} \\
& +  \sum_{k \in \mathcal{K}}\alpha_{m,k}[n] \frac{p_{m,k}[n]\omega_{m}^{\rm{U}}[n]D_{m}[n]}{R_{m,k} [n]}.
\end{split}
\end{equation}

It is not difficult to verify that $\mathcal{W}_{5}[n]$ is a convex function with respect to variables $\{f^{\rm{L}}_{m}[n], f^{\rm{U}}_{m,k}[n], f^{\rm{S}}_{m}[n]\}$. 
Furthermore, according to the following Proposition \ref{proposition1}, we can conclude that $\mathcal{W}_{6}[n]$ is a concave function with respect to variables $\{p_{m,k}[n], p_{m,\rm{LEO}}[n]\}$.

\begin{proposition}{\label{proposition1}}
 The function $\mathcal{W}_{6}[n]$ is a concave function with respect to variables $p_{m,k}[n]$, and $p_{m,\rm{LEO}}[n]$.  
\begin{proof}
Please refer to Appendix \ref{appendice1}.
\end{proof}
\end{proposition}

By introducing a set of auxiliary variables $\{ \xi_{m,k}[n],\xi_{m,{\rm{LEO}}}[n]\}$, $\mathcal{W}_{6}[n]$ can be approximated by 
\begin{equation}
\begin{split}
& \Tilde{\mathcal{W}}_{6}[n] = \frac{\ln(2) \sigma_{{\rm{LEO}}}^2 \omega_{m}^{\rm{S}}[n]D_{m}[n]}{ \left|g_{{\rm{LEO}}}\right|^{2} B_{{\rm{LEO}}} } \xi_{m,{\rm{LEO}}}[n]( e^{\frac{1}{\xi_{m,{\rm{LEO}}}[n]}} -1 ) \\
& +  \sum_{k \in \mathcal{K}}\alpha_{m,k}[n] 
\frac{ \ln(2) \sigma_{{\rm{UAV}}}^2 \omega_{m}^{\rm{U}}[n]D_{m}[n]}{  h_{m,k} [n] B_{m,k} } \xi_{m,k}[n](e^{\frac{1}{\xi_{m,k}[n]}}-1) ,
\end{split}
\end{equation}
where $\xi_{m,k}[n]$ and $\xi_{m,{\rm{LEO}}}[n]$ are given by
\begin{equation}
    \xi_{m,k}[n] = \frac{1}{\ln \left(1 +\dfrac{p_{m,k}[n] h_{m,k} [n]}{\sigma_{{\rm{UAV}}}^2} \right)},
\end{equation}
\begin{equation}
    \xi_{m,{\rm{LEO}}}[n] = \frac{1}{\ln \left(1 +\frac{p_{m,{\rm{LEO}}}[n] \left|g_{{\rm{LEO}}}\right|^{2}}{\sigma_{{\rm{LEO}}}^2} \right)}.
\end{equation}

\begin{proposition}{\label{proposition2}}
 The function $\Tilde{\mathcal{W}}_{6}[n]$ is a convex function with respect to variables $\{ \xi_{m,k}[n],\xi_{m,{\rm{LEO}}}[n]\}$.  
\begin{proof}
Please refer to Appendix \ref{appendice2}.
\end{proof}
\end{proposition}

It is worth noting that the directly expression of $\Tilde{\mathcal{W}}_{6}[n]$ in CVX is not feasible.
Therefore, we transfer it through the exponential cone optimization.
Specifically, by introducing a set of auxiliary variables $\{ \Gamma_{m,k}[n],\Gamma_{m,{\rm{LEO}}}[n]\}$, $\Tilde{\mathcal{W}}_{6}[n]$ can be rewritten as $\Tilde{\mathcal{W}}^{'}_{6}[n]$, and given by
\begin{equation}
\begin{split}
\Tilde{\mathcal{W}}^{'}_{6}[n] = & \frac{\ln(2) \sigma_{{\rm{LEO}}}^2 \omega_{m}^{\rm{S}}[n]D_{m}[n]}{ \left|g_{{\rm{LEO}}}\right|^{2} B_{{\rm{LEO}}} } \Gamma_{m,{\rm{LEO}}}[n] \\
& +  \sum_{k \in \mathcal{K}}\alpha_{m,k}[n] 
\frac{ \ln(2) \sigma_{{\rm{UAV}}}^2 \omega_{m}^{\rm{U}}[n]D_{m}[n]}{  h_{m,k} [n] B_{m,k} } \Gamma_{m,k}[n] ,
\end{split}
\end{equation}
where 
\begin{equation}
\xi_{m,{\rm{LEO}}}[n] e^{\frac{1}{\xi_{m,{\rm{LEO}}}[n]}}   \leq \Gamma_{m,{\rm{LEO}}}[n] + \xi_{m,{\rm{LEO}}}[n], \label{SP3-Constraint1}
\end{equation}
\begin{equation}
\xi_{m,k}[n] e^{\frac{1}{\xi_{m,k}[n]}} \leq \Gamma_{m,k}[n] + \xi_{m,k}[n]. 
\end{equation}

Accordingly, we can rewrite constraints (\ref{P111})-(\ref{P113}) as
\begin{equation}
    \frac{1}{\ln \left(1 +\dfrac{P_{\rm{max}}^{\rm{U}} h_{m,k} [n]}{\sigma_{{\rm{UAV}}}^2} \right)} \leq \xi_{m,k}[n] , \forall m,k,n, 
\end{equation}
\begin{equation}
    \frac{1}{\ln \left(1 +\dfrac{P_{\rm{max}}^{\rm{S}} \left|g_{{\rm{LEO}}}\right|^{2}}{\sigma_{{\rm{LEO}}}^2} \right)} \leq \xi_{m,{\rm{LEO}}}[n], \forall m,n,
\end{equation}
\begin{equation}
   L_{m}^{\rm{L}}[n]  \le \tau, \forall m,n,
\end{equation}
\begin{equation}
\begin{split}
L_{m}^{\rm{U}}[n] =& \sum_{k=1}^{K}\alpha_{m,k}[n]  \frac{ \ln (2) \omega_{m}^{\rm{U}}[n]D_{m}[n] \xi_{m,k}[n]}{ B_{m,k} } \\
& +\sum_{k=1}^{K}\alpha_{m,k}[n] \frac{ \omega_m^{\rm{U}}[n] D_{m}[n] \phi_{m}[n] }{f^{\rm{U}}_{m,k}[n]} \le \tau, \forall m,k,n,
\end{split}
\end{equation}
\begin{equation}
\begin{split}
L_{m}^{\rm{S}}[n] &= \frac{ \ln (2) \omega_{m}^{\rm{S}}[n]D_{m}[n] \xi_{m,{\rm{LEO}}}[n] }{ B_{{\rm{LEO}}} } +
\frac{ \omega_m^{\rm{S}}[n] D_{m}[n] \phi_{m}[n] }{f^{\rm{S}}_{m}[n]} \\
& \le \tau, \forall m,n. \label{SP3-Constraint2}
\end{split}
\end{equation}

To this end, the problem \textbf{SP3} is transformed into \textbf{SP3'}, which is written as
\begin{subequations}
\begin{align}
\textbf{SP3'} \ & \min\limits_{\{\boldsymbol{F},\boldsymbol{\xi},\boldsymbol{\Gamma} \}}  \sum_{m=1}^{M}\sum_{n=1}^{N}  y_{m,n}^{2} \left( \mathcal{W}_{5}[n] + \Tilde{\mathcal{W}}^{'}_{6}[n]  \right)\\
{\rm{s.t.}}\ &\text{(\ref{P18})}-\text{(\ref{P110}), (\ref{P114}), }\text{(\ref{SP3-Constraint1})}-\text{(\ref{SP3-Constraint2}),}
\end{align}
\end{subequations}
where $ \boldsymbol{\xi} = \{ \xi_{m,k}[n], \xi_{m,{\rm{LEO}}}[n], \forall m,k,n \} $, $ \boldsymbol{\Gamma} = \{ \Gamma_{m,k}[n],$
$ \Gamma_{m,{\rm{LEO}}}[n], \forall m,k,n \} $. The problem \textbf{SP3'} is a convex optimization problem, which can be solved by CVX.

\subsection{SUBPROBLEM 4: UAV TRAJECTORY OPTIMIZATION}
By fixing the variables $ \boldsymbol{P}, \boldsymbol{\Omega}, \boldsymbol{F}, \boldsymbol{\alpha} $, the subproblem to optimize the UAV trajectory is formulated as
\begin{subequations}
\begin{align}
\textbf{SP4:} \ & \min\limits_{\{\boldsymbol{Q} \}}  \sum_{m=1}^{M}\sum_{n=1}^{N}  y_{m,n}^{2} E_{m}^{\rm{U},\rm{tran}}[n] \\
{\rm{s.t.}}\ &\text{(\ref{P11})}-\text{(\ref{P13})}, \text{(\ref{P113})}-\text{(\ref{P114}).}
\end{align}
\end{subequations}

Regarding the inequality constraint (\ref{P114}), $E_{k}^{\rm{prop}}[n]$ contains $\frac{\zeta_2}{\|\boldsymbol{v}_{k}[n]\|}$, which is a reciprocal of a convex function.
Therefore, the constraint (\ref{P114}) is a non-convex constraint. To address this, we introduce an auxiliary variable $\psi_{k}[n]$, which satisfies
\begin{equation}
 \psi_{k}^2[n] \le \frac{\|\boldsymbol{q}_{k}[n]-\boldsymbol{q}_{k}[n-1]\|^2}{\tau^2 } . \label{SP4-1}
\end{equation}

Therefore, $E_{k}^{\rm{prop}}[n]$ can be approximated by its convex upper bound
\begin{equation}
\hat{E}_{k}^{\rm{prop}}[n] = \tau \left(\zeta_1\|\boldsymbol{v}_{k}[n]\|^3+\frac{\zeta_2}{{\psi}_{k}[n]} \right).
\end{equation}

By replacing $E_{k}^{\rm{prop}}[n]$ with $\hat{E}_{k}^{\rm{prop}}[n]$, the inequality constraint (\ref{P114}) can be transformed into
\begin{equation}
\sum_{n \in \mathcal{N}} \left( \hat{E}_{k}^{\rm{prop}}[n] + \sum_{m \in \mathcal{M}}E_{m,k}^{\rm{U},\rm{com}}[n] \right) \le E_{\rm{max}}^{{\rm{UAV}}}. \label{SP4-2}
\end{equation}

Moreover, the constraint (\ref{SP4-1}) is non-convex. We convert it into the following convex form via the successive convex approximation (SCA) method \cite{9273074, 10606316}, which is given by
\begin{equation}
	\begin{split}
		& \psi_{k}^2[n]\tau^2 \le \|\boldsymbol{q}_{k}^{(l)}-\boldsymbol{q}_{k}^{(l)}[n-1]\|^2\\
		&+2(\boldsymbol{q}_{k}^{(l)}[n]-\boldsymbol{q}_{k}^{(l)}[n-1])^{\mathsf{T}}(\boldsymbol{q}_{k}[n]-\boldsymbol{q}_{k}[n-1]). \label{SP4-3}
	\end{split}
\end{equation}
Similarly, the constraint (\ref{P12}) can be transformed as
\begin{equation}
	\begin{split}
            d_{\min}^{2} \le & -\| \boldsymbol{q}^{(l)}_k[n] -\boldsymbol{q}^{(l)}_i[n] \|^{2}\\
		&+2(\boldsymbol{q}^{(l)}_k[n] -\boldsymbol{q}^{(l)}_i[n])^{\mathsf{T}}(\boldsymbol{q}_k[n] -\boldsymbol{q}_i[n]). \label{SP4-4}
	\end{split}
\end{equation}

As for constraint (\ref{P113}), only $L_{m}^{\rm{U}}[n]$ is related to the UAV trajectory. Thus, we mainly consider the equivalent constraint $L_{m}^{\rm{U}}[n] \le \tau $. By introducing auxiliary variables $S_{m,k}[n] $ and $\gamma_{m,k}[n] = R_{m,k}[n]$, where $S_{m,k}[n] $ and $\gamma_{m,k}[n]$ satisfy
\begin{equation}
    S_{m,k}[n] \le H^{2} + \| \boldsymbol{q}_k[n] - \boldsymbol{s}_m[n] \|^{2},
\end{equation}
\begin{equation}
    \gamma_{m,k}[n] \le  B_{m,k} \log_2 \left( 1 + \frac{p_{m,k}[n] \beta_{0}}{S_{m,k}[n] \sigma_{{\rm{UAV}}}^2} \right), 
\end{equation}
$L_{m}^{\rm{U}}[n] \le \tau $ can be transformed as 
\begin{equation}
\sum_{k=1}^{K}\alpha_{m,k}[n] \left( \frac{\omega_{m}^{\rm{U}}[n]D_{m}[n]}{ \gamma_{m,k}[n] } +
\frac{ \omega_m^{\rm{U}}[n] D_{m}[n] \phi_{m}[n] }{f^{\rm{U}}_{m,k}[n]} \right) \le \tau.\label{SP4-5}
\end{equation}
It is easy to prove that $B_{m,k} \log_2 \left( 1 + \frac{p_{m,k}[n] \beta_{0}}{S_{m,k}[n] \sigma_{{\rm{UAV}}}^2} \right)$ is convex with respect to $S_{m,k}[n]$. 
Then, similar to the transformation in \eqref{SP4-3}, the constraint (\ref{SP4-5}) can be converted into
\begin{equation}
\begin{split}
&\gamma_{m,k}[n] \le  B_{m,k} \Bigg( \log_2 \left( S_{m,k}[n] \sigma_{{\rm{UAV}}}^2 + p_{m,k}[n] \beta_{0} \right) \\
&- \log_{2}(S^{l}_{m,k}[n] \sigma_{{\rm{UAV}}}^2) -\frac{\log_{2}(e)}{S^{l}_{m,k}[n]}\left( S_{m,k}[n]- S^{l}_{m,k}[n] \right) \Bigg). \label{SP4-6}
\end{split}
\end{equation}
Similarly, the constraint $S_{m,k}[n] \le H^{2} + \| \boldsymbol{q}_k[n] - \boldsymbol{s}_m[n] \|^{2}$ can be transformed as
\begin{equation}
\begin{split}
S_{m,k}[n] \le & \| \boldsymbol{q}^{l}_k[n] - \boldsymbol{s}_m[n]  \|^{2} + H^2 \\
& + 2(\boldsymbol{q}^{l}_k[n] - \boldsymbol{s}_m[n])^{\mathsf{T}}(\boldsymbol{q}_k[n] - \boldsymbol{q}^{l}_k[n]). \label{SP4-7}
\end{split}
\end{equation}
Then, the objective function is written as
\begin{equation}
\begin{split}
        \Upsilon_{m}[n] &=  y_{m,n}^{2} E_{m}^{\rm{U},\rm{tran}}[n]\\
        &=y_{m,n}^{2} \sum_{k=1}^{K}\alpha_{m,k}[n] \frac{p_{m,k}[n]\omega_{m}^{\rm{U}}[n]D_{m}[n]}{ \gamma_{m,k}[n]}.
\end{split}
\end{equation}
To this end, the problem \textbf{SP4} is transformed into the following \textbf{SP4'}, which is written as
\begin{subequations}
\begin{align}
\textbf{SP4':} \ & \min\limits_{\{\boldsymbol{Q},\boldsymbol{\psi},\boldsymbol{\gamma},\boldsymbol{S}  \}}  \sum_{m=1}^{M}\sum_{n=1}^{N}  \Upsilon_{m}[n] \\
{\rm{s.t.}}\ &\text{\eqref{P11}, \eqref{P13}, \eqref{SP4-2}}-\text{\eqref{SP4-4}, \eqref{SP4-5}}-\text{\eqref{SP4-7},}
\end{align}
\end{subequations}
where $\boldsymbol{\psi} = \{\psi_{k}[n],\forall k,n \}$, $\boldsymbol{\gamma} = \{ \gamma_{m,k}[n] ,\forall m,k,n \}$, 
$\boldsymbol{S} = \{S_{m,k}[n],\forall m,k,n  \}$.
The problem \textbf{SP4'} is a convex optimization problem, which can be solved by CVX.

\subsection{ALGORITHM ANALYSIS}
The AO-based algorithm for
solving problem \textbf{P1}
is detailed in Algorithm \ref{Algorithm1}, which alternately optimizes the MU-UAV association, UAV trajectory, task offloading decision, and transmit power control and CPU frequency control in an iterative manner until the objective value converges or the maximum iteration number is reached.

\begin{algorithm} \footnotesize
\caption{AO-based Algorithm for Solving \textbf{P1}}
\label{Algorithm1}
\begin{algorithmic}
\REQUIRE {An initial feasible point $\{\boldsymbol{Q}^{0}, \boldsymbol{P}^{0}, \boldsymbol{\Omega}^{0}, \boldsymbol{F}^{0}, \boldsymbol{\alpha}^{0} \}$;}\\
\textbf{Initialize:} Iteration number $l=0$, precision threshold $\varepsilon$, and number of maximum iterations $l_{\max}$;\\
\STATE Compute the initial quadratic transform variable $y_{m,n}^{0}$ according to (\ref{P21}) and the initial objective function value, i.e. $\Phi^{0}\left(\boldsymbol{Q}^{0}, \boldsymbol{P}^{0}, \boldsymbol{\Omega}^{0}, \boldsymbol{F}^{0}, \boldsymbol{\alpha}^{0}\right)$;
\REPEAT
\STATE Solve the problem \textbf{SP1} to get the MU-UAV association $\boldsymbol{\alpha}^{l+1}$ for given $\boldsymbol{Q}^{l}, \boldsymbol{P}^{l}, \boldsymbol{\Omega}^{l}, \boldsymbol{F}^{l} $;\\
\STATE Solve the problem \textbf{SP2} to get the task offloading decosion $\boldsymbol{\Omega}^{l+1}$ for given $\boldsymbol{Q}^{l}, \boldsymbol{P}^{l}, \boldsymbol{F}^{l}, \boldsymbol{\alpha}^{l+1} $;\\
\STATE Solve the problem \textbf{SP3'} to get the transmit power control and CPU frequency control $\boldsymbol{P}^{l+1}, \boldsymbol{F}^{l+1}$ for given $\boldsymbol{Q}^{l},  \boldsymbol{\Omega}^{l+1}, \boldsymbol{\alpha}^{l+1} $;\\
\STATE Solve the problem \textbf{SP4'} to get the UAV trajectory $\boldsymbol{Q}^{l+1}$ for given $\boldsymbol{P}^{l+1}, \boldsymbol{\Omega}^{l+1}, \boldsymbol{F}^{l+1}, \boldsymbol{\alpha}^{l+1}$;\\
\STATE Update the quadratic transform variable $y_{m,n}^{l+1}$ according to (\ref{P21});
\STATE Update the objective function value according to above variables, i.e. $\Phi\left(\boldsymbol{Q}^{l+1}, \boldsymbol{P}^{l+1}, \boldsymbol{\Omega}^{l+1}, \boldsymbol{F}^{l+1}, \boldsymbol{\alpha}^{l+1}\right)$;
\STATE Update $l=l+1$; \\
\UNTIL The objective function between two adjacent iterations is smaller than precision threshold $\varepsilon$ or $l > l_{\max}$;\\
\ENSURE {$\Phi^{*}\left(\boldsymbol{Q}^{*}, \boldsymbol{P}^{*}, \boldsymbol{\Omega}^{*}, \boldsymbol{F}^{*}, \boldsymbol{\alpha}^{*}\right)$, $\boldsymbol{Q}^{*}$, $\boldsymbol{P}^{*}$, $\boldsymbol{\Omega}^{*}$, $\boldsymbol{F}^{*}$,  and $\boldsymbol{\alpha}^{*}$}.\\
\end{algorithmic}
\end{algorithm}

\subsubsection{Convergence Anlysis}
Similar to the notations in Algorithm \ref{Algorithm1},
we define $\boldsymbol{Q}^{l}$, $\boldsymbol{P}^{l}$, $\boldsymbol{\Omega}^{l}$, $\boldsymbol{F}^{l}$, and $\boldsymbol{\alpha}^{l}$ as the solution in the $l$th iteration. The objective function is defined as $\Phi\left(\boldsymbol{Q}^{l}, \boldsymbol{P}^{l}, \boldsymbol{\Omega}^{l}, \boldsymbol{F}^{l}, \boldsymbol{\alpha}^{l}\right)$.
Thus, we have
\begin{equation}
\begin{split}
\Phi\left(\boldsymbol{Q}^{l}, \boldsymbol{P}^{l}, \boldsymbol{\Omega}^{l}\right.,& \left.\boldsymbol{F}^{l}, \boldsymbol{\alpha}^{l}\right) \leq \Phi\left(\boldsymbol{Q}^{l}, \boldsymbol{P}^{l}, \boldsymbol{\Omega}^{l}, \boldsymbol{F}^{l}, \boldsymbol{\alpha}^{l+1}\right) \\
&\leq \Phi\left(\boldsymbol{Q}^{l}, \boldsymbol{P}^{l}, \boldsymbol{\Omega}^{l+1}, \boldsymbol{F}^{l}, \boldsymbol{\alpha}^{l+1}\right)\\
&\leq \Phi\left(\boldsymbol{Q}^{l}, \boldsymbol{P}^{l+1}, \boldsymbol{\Omega}^{l+1}, \boldsymbol{F}^{l+1}, \boldsymbol{\alpha}^{l+1}\right)\\
&\leq \Phi\left(\boldsymbol{Q}^{l+1}, \boldsymbol{P}^{l+1}, \boldsymbol{\Omega}^{l+1}, \boldsymbol{F}^{l+1}, \boldsymbol{\alpha}^{l+1}\right),
\end{split}
\end{equation}
which shows that the value of the objective function is non-decreasing over iterations. In addition, the energy efficiency is bounded due to the finite range of the optimization variables. Therefore, Algorithm \ref{Algorithm1} is guaranteed to converge.

\subsubsection{Complexity Anlysis}
In Algorithm \ref{Algorithm1}, we solve the subproblem \textbf{SP1} using the binary cut-and-branch method \cite{2013MunozJoint}, and solve other subproblems using the interior point method \cite{2010LuoSemidefinite}. 
Thus, the computational complexity of the four subproblems are $\mathcal{O}\left(MKN\log(MKN)\right)$, $\mathcal{O}\left((MN)^{3.5}\log(1/\epsilon)\right)$, $\mathcal{O}\left((MKN)^{3.5}\log(1/\epsilon)\right)$, $\mathcal{O}\left((MKN)^{3.5}\log(1/\epsilon)\right)$, respectively, where $\epsilon$ represents the tolerance. Consequently, the complexity of Algorithm \ref{Algorithm1} is 
$\mathcal{O}( \mathcal{I} (((MN)^{3.5} $
$+ (MKN)^{3.5} + (MKN)^{3.5})\log(1/\epsilon) + MKN\log(MKN) ) )$,
where $\mathcal{I}$ denotes the iteration number.

\section{NUMERICAL RESULTS}\label{section5}
In the simulations, we consider $K=2$ UAVs fly from specified initial positions to specified final positions to provide computing services for the MUs. We consider $M=8$ MUs that are randomly distributed in a square area of $1000 \times 1000 \ \rm{m}^{2}$. All UAVs fly at a height of 100 $\rm{m}$. Unless otherwise stated, the remaining simulation parameters are summarized in Table \ref{table1} \cite{Chen2022Energy}.

\begin{table}[!htbp] \footnotesize
\caption{Simulation Parameters.}
\centering
\label{table1}
\begin{tabular}{p{6.1cm}c}
\toprule
\centering
Parameters & Value\\
\midrule  %
\centering Task period, $T$ &  40 s\\
\centering Number of time slot, $N$ & 40\\
\centering Duration of time slot, $\tau$ & 1 s\\
\centering Minimum secure distance, $d_{\rm{min}}$& 50 m\\
\centering Maximum speed, $V_{\rm{max}}$& 50 m/s\\
\centering UAV’s propulsion parameters, $\zeta_{1},\zeta_{2}$& 0.00614,15.976\\
\centering Channel power gain between the UAVs and the MUs, $\beta_{0}$& -60 dBm\\
\centering Noise power at the UAV, $\sigma_{\rm{UAV}}$& -170 dBm\\
\centering Channel power gain between the LEO satellite and the MUs, $g_{\rm{LEO}}$& [5,10] dBm \\
\centering Noise power at the LEO satellite, $\sigma_{\rm{LEO}}$& -170 dBm\\
\centering Maximum transmit power, $P_{\rm{max}}$ & 3 w\\
\centering Data size, $D_{m}[n]$& [5.5, 6.5] Mbits\\
\centering Computing cycles for 1-bit data, $\phi_{m}[n]$ & 100 cycles/bit\\
\centering MU maximum computing frequency, $F_{\rm{max}}^{\rm{L}}$ & 5 $\rm{Ghz}$\\
\centering MU computing capacitance coefficient, $\kappa_{\rm{L}}$ & $10^{-26}$\\
\centering UAV maximum computing frequency, $F_{\rm{max}}^{\rm{U}}$& 9 $\rm{Ghz}$\\
\centering UAV computing capacitance coefficient, $\kappa_{\rm{U}}$& $10^{-27}$\\
\centering LEO satellite maximum computing frequency, $F_{\rm{max}}^{\rm{S}}$& 9 $\rm{Ghz}$\\
\centering LEO satellite computing capacitance coefficient, $\kappa_{\rm{S}}$& $10^{-27}$\\
\bottomrule
\end{tabular}
\end{table}

To evaluate the performance and efficiency of the proposed
scheme, we compare it with four baseline schemes, which are given in the following.
\begin{itemize}
\item[1)] \emph{Single UAV Scheme:} The considered SAGIN system has only one UAV that can provide computing services for the MUs.
\item[2)] \emph{Fixed Trajectory Scheme:} The trajectories of the UAVs are fixed.
\item[3)] \emph{Fixed Allocation Scheme:} The task offloading decisions of the MUs are fixed.
\item[4)] \emph{LS Offloading Scheme:} Only the LEO satellite provides computing services for the MUs.
\end{itemize}

\begin{figure}[!htbp]
	\centering
	\includegraphics[width=0.88\linewidth]{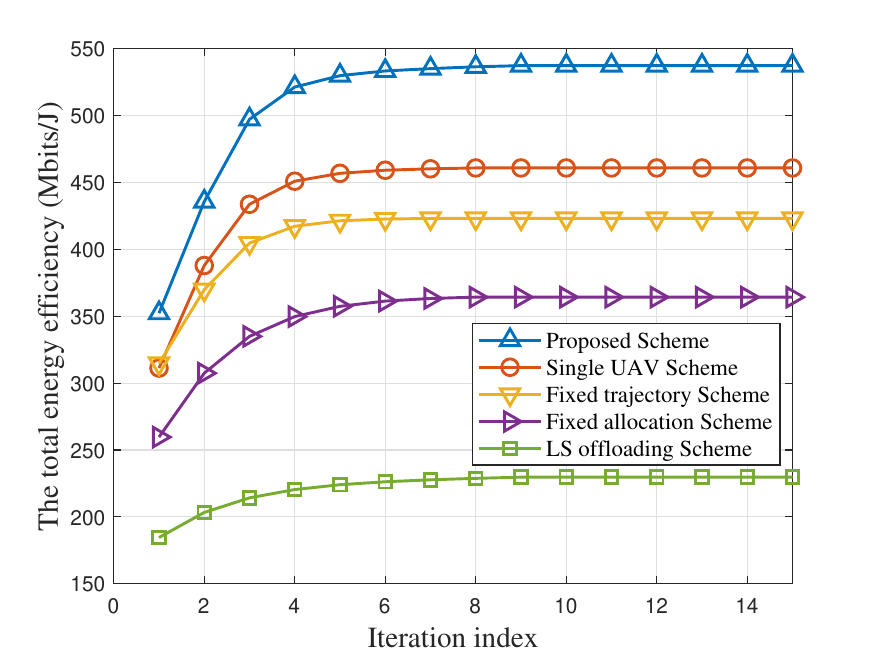}
	\caption{The total energy efficiency under different schemes versus the number of iterations.}
	\label{convergence}
\end{figure}

As shown in Fig. \ref{convergence}, we present the total energy effi-
ciency under different schemes with respect to the number of iterations. The results clearly show that the total energy efficiency initially exhibits a significant increase, followed by a gradual convergence to a stable value as the number of iterations progresses, which validates the convergence of all schemes. Notably, the total energy efficiency of the proposed scheme is the highest among all schemes, which can be attributed to the joint design of the MU-UAV association, UAV trajectory, task offloading decision, transmit power control, and CPU frequency control.

\begin{figure}[!htbp]
	\centering
	\includegraphics[width=0.88\linewidth]{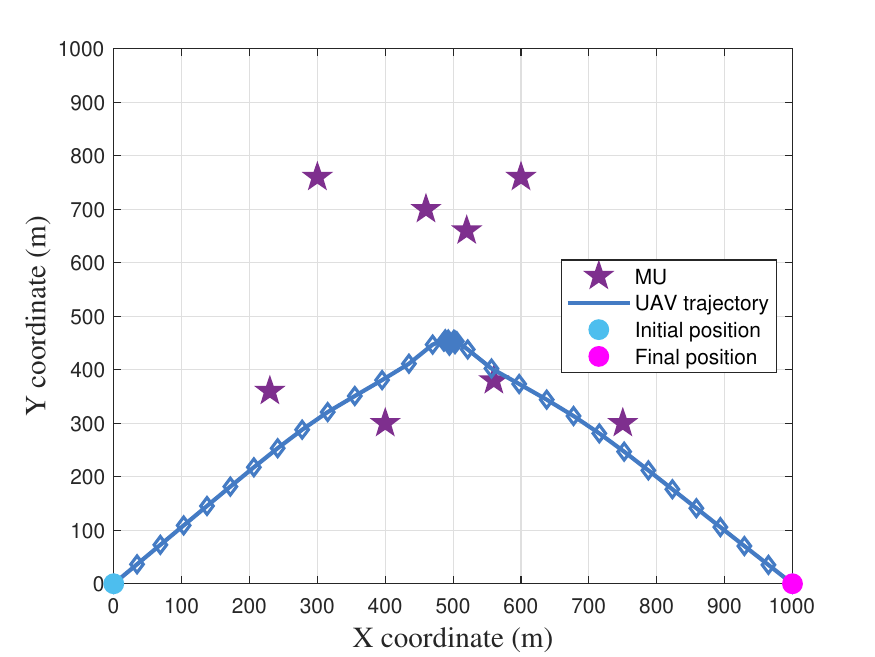}
	\caption{The optimized trajectory of the UAV under the single UAV scheme.}
	\label{uav trajectory1}
\end{figure}

\begin{figure}[!htbp]
	\centering
	\includegraphics[width=0.88\linewidth]{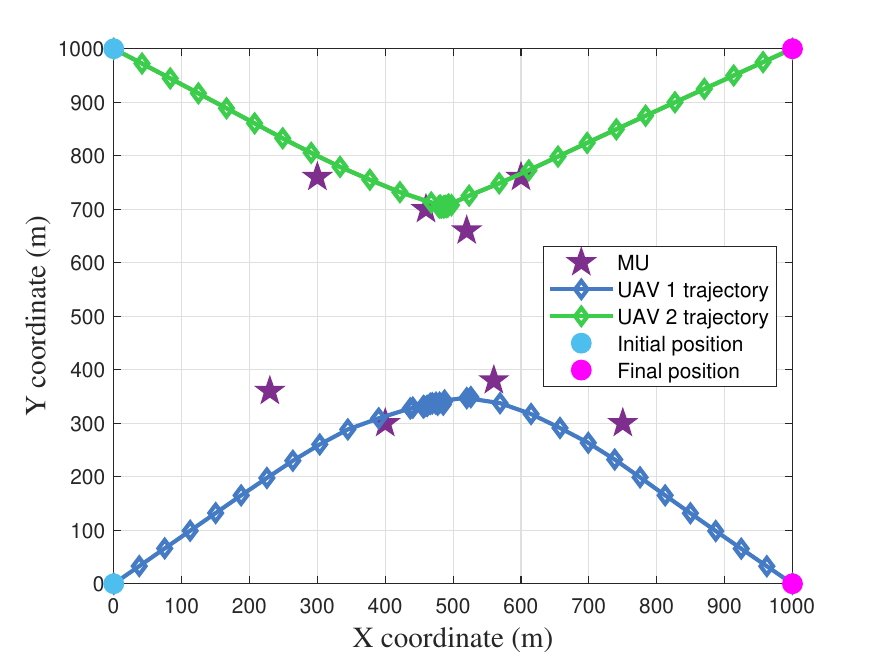}
	\caption{The optimized trajectories of the UAVs under the proposed scheme.}
	\label{uav trajectory2}
\end{figure}

Fig. \ref{uav trajectory1} and Fig. \ref{uav trajectory2} respectively illustrate the optimized trajectories of the UAVs under the single UAV scheme and the proposed scheme. It can be observed that the UAV trajectories consistently tend towards regions densely populated with the MUs. This proximity typically enhances the transmission rates between the MUs and the UAVs, thereby improving the timelines of service responses. Further, the comparative analysis of Fig. \ref{uav trajectory1} and Fig. \ref{uav trajectory2} reveals that as the number of the UAVs increases, the number of the MUs served by each UAV decreases. This enables the UAVs to more effectively optimize their positions to approach the MUs they serve, leading to a substantial improvement in the transmission rates between the UAVs and the MUs they serve.

\begin{figure}[!htbp]
	\centering
	\includegraphics[width=0.88\linewidth]{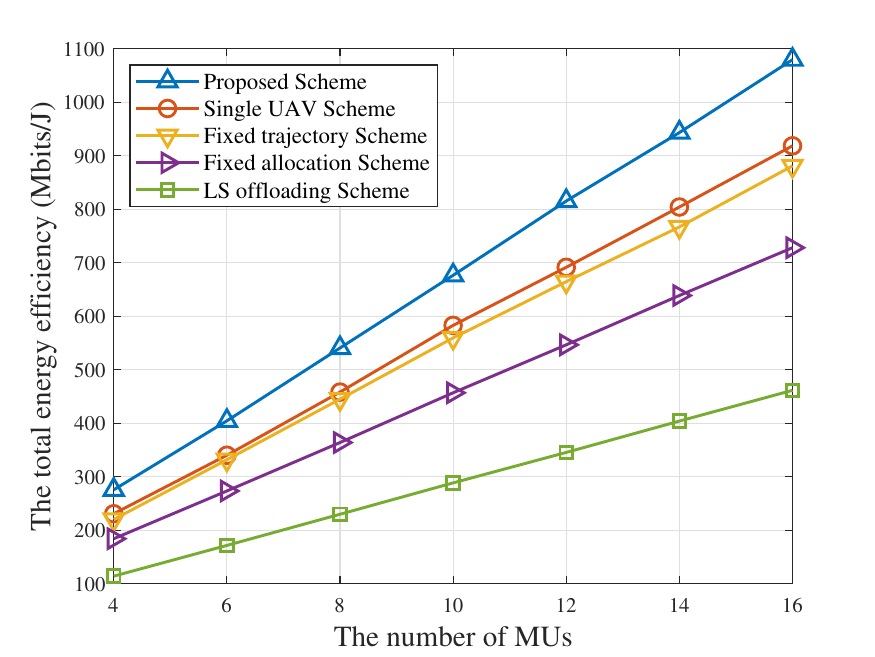}
	\caption{The total energy efficiency versus the number of MUs.}
	\label{MU number1}
\end{figure}

\begin{figure}[!htbp]
	\centering
	\includegraphics[width=0.88\linewidth]{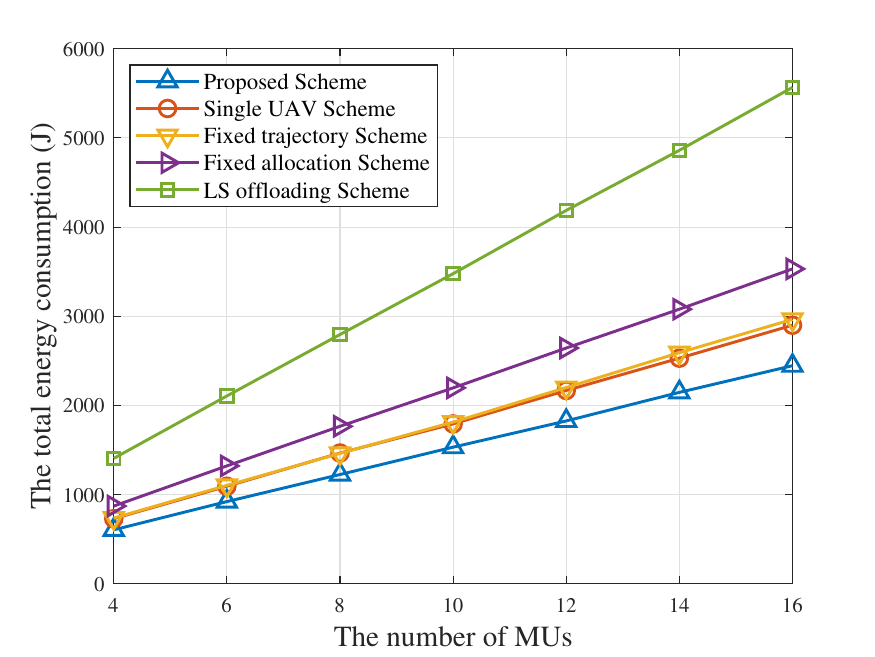}
	\caption{The total energy consumption versus the number of MUs.}
	\label{MU number2}
\end{figure}

Fig. \ref{MU number1} and Fig. \ref{MU number2} show that, as the number of MUs increases, our proposed scheme exhibits a more significant increase in total energy efficiency compared to other four schemes, while the increase in energy consumption is relatively smaller.
The reason can be explained as follows. As the number of MUs grows, the demand for the UAV services increases, and the various schemes exhibit different levels of adaptability to these changes.
Specifically, the single UAV scheme deploys only a single UAV, which makes it challenging to ensure high-quality communication services for every MU when a large number of MUs require for the services. 
The fixed trajectory scheme lacks flexibility, especially when there are multiple UAVs. This means that in scenarios with the uneven MU distribution or dynamic demand changes, the scheme struggles to optimize service quality by adjusting the UAV positions.
The fixed allocation scheme employs a fixed task allocation decision, which means it cannot adjust the allocation of tasks based on real-time network conditions. In contrast, our scheme dynamically reallocates tasks according to the current network state, ensuring higher service quality and better user experience.
The LS offloading scheme relies solely on local computing resources and the LEO satellite assisted computing without considering the role of the UAVs. Although this approach can provide a certain level of service, the absence of the UAV support makes it difficult to achieve ideal computational quality and efficiency in scenarios with a large-scale MU distribution.
Therefore, under the same number of MUs, our proposed scheme demonstrates the best performance. Specifically, from the numerical results, when the number of the MUs is $16$, our proposed scheme outperforms the single UAV scheme, fixed trajectory scheme, fixed allocation scheme, and LS offloading scheme by $17.53\%$, $22.55\%$, $48.27\%$, and $133.90\%$, respectively, in terms of the overall computational energy efficiency. 

\begin{figure}[!htbp]
	\centering
	\includegraphics[width=0.88\linewidth]{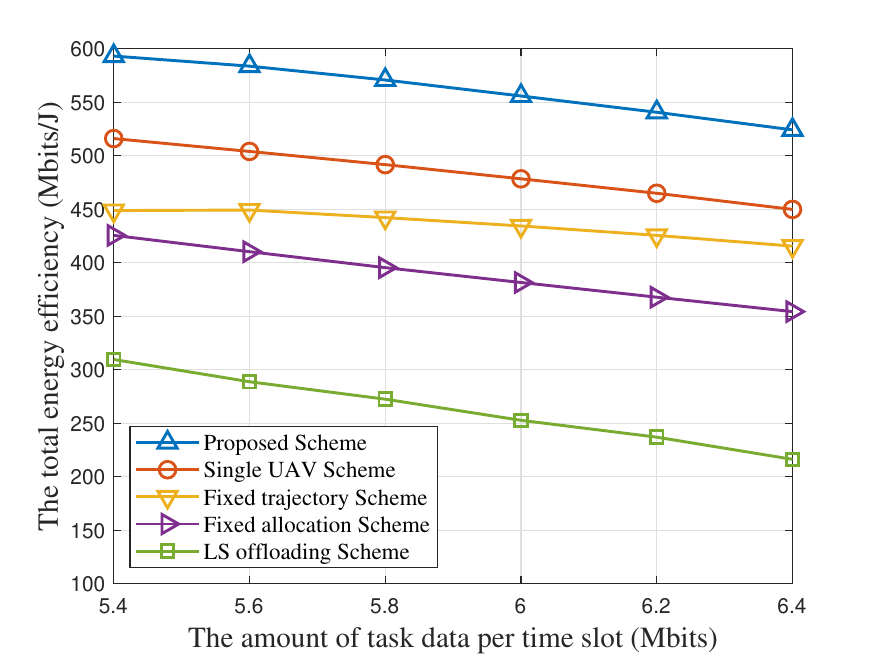}
	\caption{The total energy efficiency versus the amount of task data per time slot.}
	\label{task size1}
\end{figure}

\begin{figure}[!htbp]
	\centering
	\includegraphics[width=0.88\linewidth]{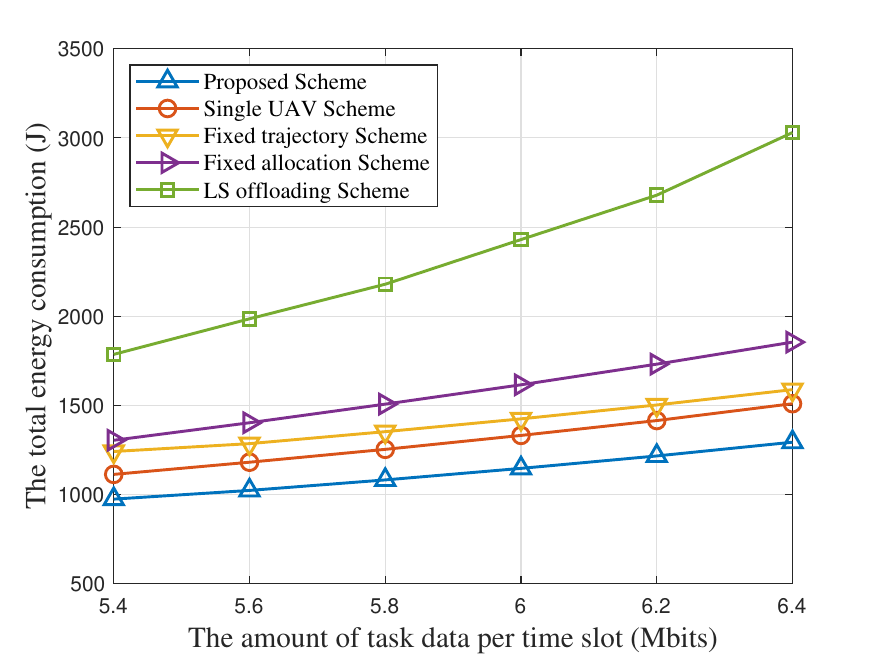}
	\caption{The total energy consumption versus the amount of task data per time slot.}
	\label{task size2}
\end{figure}

Fig. \ref{task size1} and Fig. \ref{task size2} show that, as the amount of task data that the MUs need to process per time slot increases, the total energy efficiency of all schemes decreases, and the required energy consumption also increases. Notably, in both key metrics of total energy efficiency and energy consumption, our proposed algorithm outperforms other four algorithms.
The reason is that as the amount of task data increases, the MUs tend to offload more tasks to the UAVs and the LEO satellite for remote computing, leading to an increase in communication energy consumption, which in turn reduces the total energy efficiency of all schemes. Moreover, the LS offloading scheme experiences the most significant increase in total energy consumption, and the rate of increase is notably faster than that of other four schemes. The reason is that, in the LS offloading scheme, the MUs do not offload tasks to the UAVs.

\begin{figure}[!htbp]
	\centering
	\includegraphics[width=0.88\linewidth]{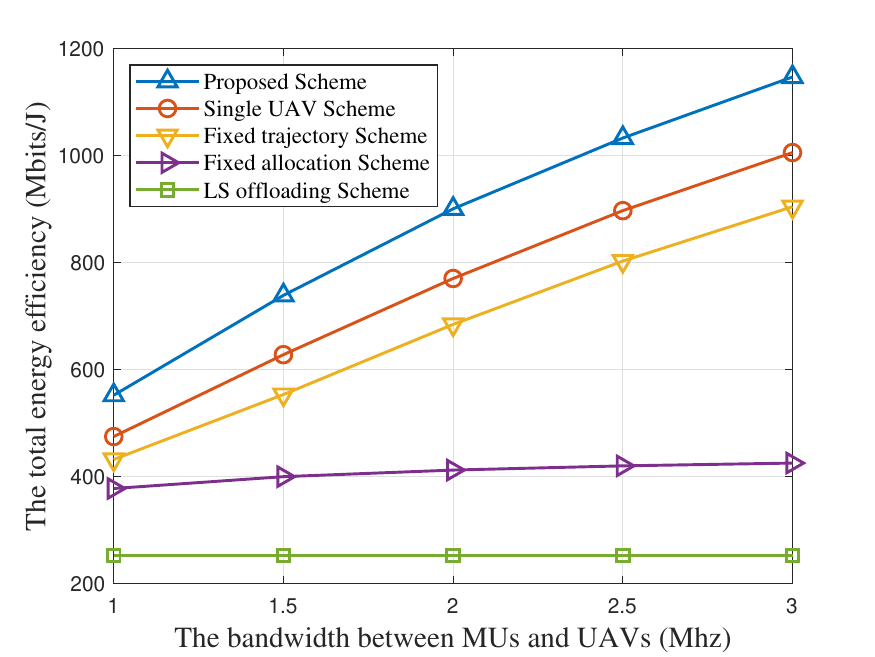}
	\caption{The total computational energy efficiency versus the bandwidth of UAVs.}
	\label{bandwidth1}
\end{figure}

Fig. \ref{bandwidth1} presents the total computational energy efficiency versus the communication bandwidth between the UAVs and the MUs. It can be observed that the total energy efficiency significantly increases with the growth of the communication bandwidth between the MUs and the UAVs for the proposed scheme, single UAV scheme, and fixed trajectory scheme, whereas the increase is relatively slower for fixed allocation scheme, and remains unchanged for LS offloading scheme. The reason is that the increase in communication bandwidth enhances the task transmission rate between the MUs and the UAVs, allowing the UAVs to assist in processing a larger volume of tasks within each time slot $\tau$.
Specifically, for the proposed scheme, single UAV scheme, and fixed trajectory scheme, the increase in communication bandwidth means that the MUs can offload tasks to the UAVs more quickly, improving the efficiency of task processing and reducing the burden of local computation, thus enhancing overall energy efficiency. However, in the fixed allocation scheme, since the proportions of task offloading to the UAVs is fixed, the growth in communication bandwidth can only reduce the energy consumption associated with the UAV computing, but does not enable the MUs to offload more tasks to the UAV, thereby limiting its potential for improvement in energy efficiency.
For the LS offloading scheme, since there is no UAV involved, the increase in communication bandwidth has no effect, and the total energy efficiency remains a constant.

\begin{figure}[!htbp]
	\centering
	\includegraphics[width=0.88\linewidth]{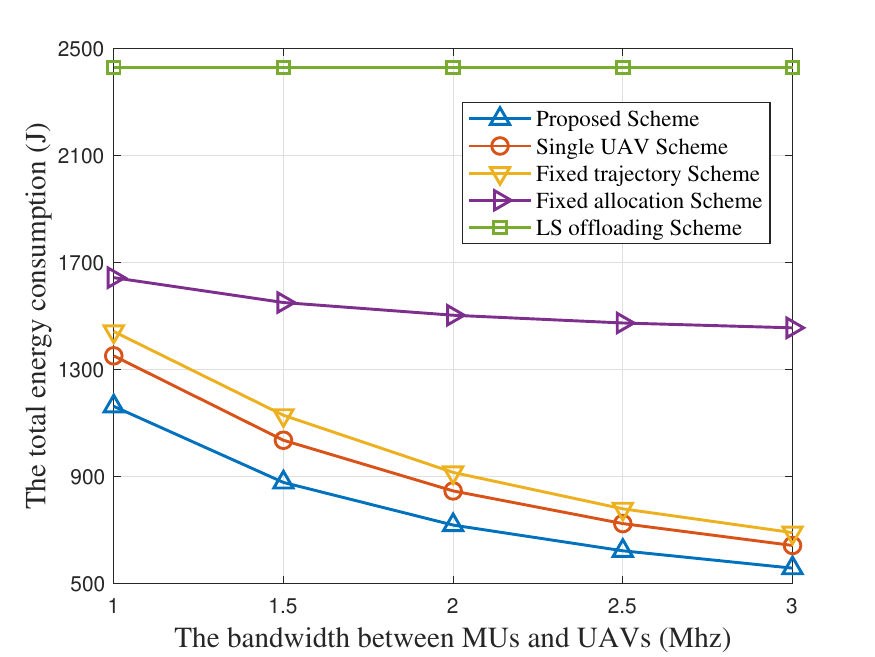}
	\caption{The total energy consumption versus the bandwidth of UAVs.}
	\label{bandwidth2}
\end{figure}

\begin{figure}[!htbp]
	\centering
	\includegraphics[width=0.88\linewidth]{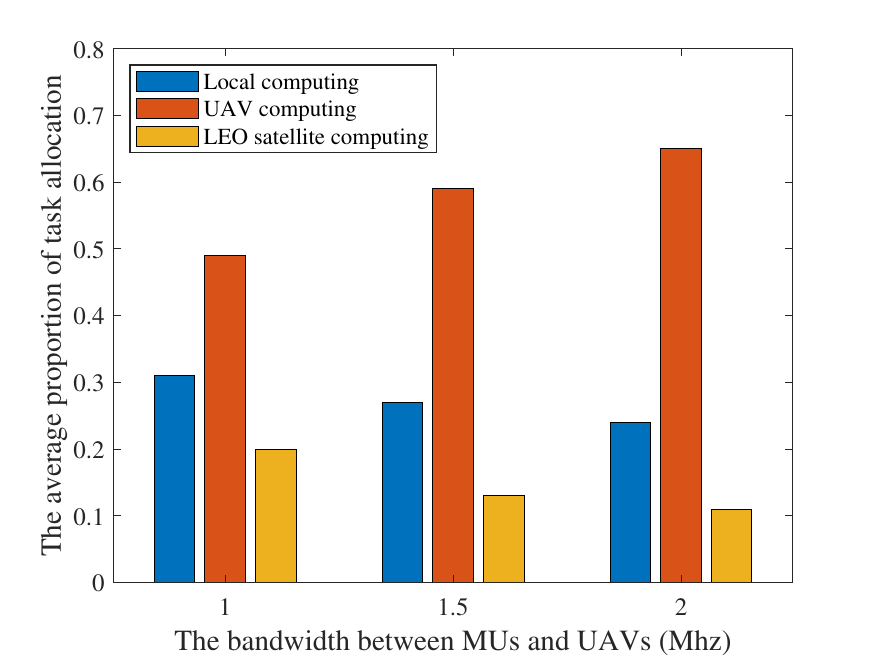}
	\caption{Task allocation at different communication bandwidth.}
	\label{bandwidth3}
\end{figure}

Fig. \ref{bandwidth2} presents the total energy consumption versus the communication bandwidth between the UAVs and the MUs.
The proposed scheme, single UAV scheme, and fixed trajectory scheme demonstrate a significant reduction in total energy consumption as the communication bandwidth increases, whereas the reduction is relatively slower for the fixed allocation scheme, and remains unchanged for the LS offloading scheme. The reason is that the increase in communication bandwidth significantly enhances the communication rate between the MUs and the UAVs, thereby increasing the proportion of tasks offloaded from the MUs to the UAVs, which is shown in Fig. \ref{bandwidth3}. Specifically, when the communication bandwidth increases from $1$ MHz to $2$ MHz, the average proportion of tasks offloaded to the UAVs rises from $49\%$ to $65\%$, while the average proportion of tasks processed locally decreases from $31\%$ to $24\%$ in the proposed scheme. Since the MUs are equipped with low energy-efficient processors, offloading more tasks to the UAVs for processing can effectively lower the total energy consumption of the system.

\begin{figure}[!htbp]
	\centering
	\includegraphics[width=0.88\linewidth]{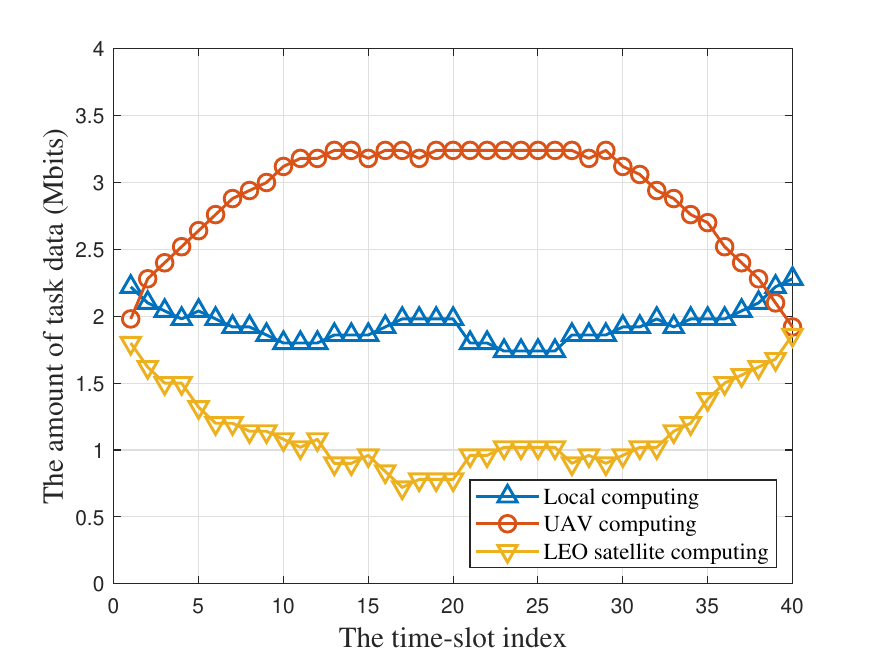}
	\caption{Task allocation at each slot.}
	\label{allocation}
\end{figure}

From Fig. \ref{allocation}, we can observe that the amount of task data offloaded to the UAVs initially increases gradually, then stabilizes, and finally decreases as time elapses. This phenomenon is attributed to the channel gains between the UAVs and the MUs. In the initial time slot, the UAVs move towards the area with higher MU density. During this movement, as the UAVs approach the MUs, the communication rates between the MUs and the UAVs increase, leading to a corresponding increase in the size of tasks offloaded to the UAVs.
Between the time slots $11$ and $29$, the amount of task data offloaded to the UAVs generally remains a constant. The reason is that during this period, the UAVs are hovering near the area with high MU density, maintaining a high communication rate, which keeps the level of task offloading stable. As the index of the time slot continues to increase, the UAVs move near to the final position, causing a decrease in the communication rates between the UAVs and the MUs, which in turn reduces the amount of task data offloaded to the UAVs.
Meanwhile, the amount of task data processed locally by the MUs and the amount of task data offloaded to the LEO satellite exhibit a divergent trend from that of the UAVs. This trend reflects adaptive task offloading decisions in response to changes in the UAVs' position and communication conditions, aiming to maximize the overall energy efficiency.

\section{CONCLUSION}\label{section6}
In this paper, we design a multi-UAV enabled SAGIN with MEC, aiming to provide efficient computing services for the MUs. Specifically, under the constraints of task delay and UAV energy limitations, we formulate a joint optimization problem targeting the maximization of the total energy efficiency. The problem involves optimizing the UAV trajectories, the MU-UAV association, the task offloading decision, the computing frequency, and the transmission power control.
To solve the problem, we decompose it into four subproblems and propose an effective AO-based algorithm. Additionally, we conduct detailed analysis of the convergence and computational complexity of the proposed algorithm.
Numerical results demonstrate that
our proposed scheme significantly outperforms other benchmark schemes. These results confirm that our scheme not only effectively improves the computational energy efficiency of the system, but also meets the computational demands of the MUs, particularly in scenarios with high number of MUs.

% \section*{APPENDIX A}

\begin{appendices}
\section{ The proof of proposition \ref{proposition1} }\label{appendice1}
\begin{proof}
Define a function $F(x) = \frac{Ax}{\ln(1+Bx)}$, with $x \ge 0$, and $A > 0$, $B > 0$ are two constants. First, we prove the non-convexity of the function $F(x)$.
The second-order derivatives of $F(x)$ with regard to $x$ is
\begin{equation}
\begin{split}
    \frac{d^2F(x)}{dx^2} =  \frac{ 2AB^{2}x - (AB^{2}x+2AB)\ln(1+Bx)    
    }{(1+Bx)^{2}\ln^{3}(1+Bx)}.
\end{split}
\end{equation}

Let $G(x) = 2AB^{2}x - (AB^{2}x+2AB)\ln(1+Bx)$, the first-order derivative and second-order derivative of $G(x)$ with regard to $x$ are
\begin{equation}
 \frac{dG(x)}{dx} = \frac{AB^{3}x}{1+Bx} - AB^{2}\ln(1+Bx),
\end{equation}
\begin{equation}
 \frac{d^2G(x)}{dx^2} = -\frac{AB^{4}x}{(1+Bx)^{2}}.
\end{equation}

Clearly, $\frac{d^2G(x)}{dx^2} \le 0$. Thus, $\frac{dG(x)}{dx}$ is a monotonically non-increasing function, and $\frac{dG(x)}{dx} \le \frac{dG(x)}{dx}|_{x=0}=0$. Consequently, $G(x)\le G(0)=0$. Thus, $\frac{d^2F(x)}{dx^2} \le 0$, indicating that $F(x)$ is a concave function with respect to $x$.

Let $A$, $B$ be $ \sum_{k \in \mathcal{K}}\alpha_{m,k}[n] \frac{\omega_{m}^{\rm{U}}[n]D_{m}[n]\ln(2)}{ B_{m,k}}$, $\frac{h_{m,k} [n]}{\sigma_{{\rm{UAV}}}^2}$, respectively. By replacing $x$ with $p_{m,k}[n]$, we conclude that the function $\mathcal{W}_{6}[n]$ is non-convex with respect to $p_{m,k}[n]$. Similarly, we can also conclude that the function $\mathcal{W}_{6}[n]$ is concave with respect to $p_{m,\rm{LEO}}[n]$. This proof is completed.
\end{proof}

\section{ The proof of proposition \ref{proposition2} }\label{appendice2}
\begin{proof}
Define $H(x) = Ax(e^{\frac{1}{x}}-1)$, with $x > 0$, and $A > 0$ is constants.
Since
\begin{equation}
\begin{split}
    \frac{d^2 H(x)}{dx^2} = \frac{e^{\frac{1}{x}}}{x^{3}} > 0,
\end{split} 
\end{equation}
one can get that $H(x)$ is a convex function with respect to $x$.

Let $A = \alpha_{m,k}[n] 
\frac{ \ln(2) \sigma_{{\rm{UAV}}}^2 \omega_{m}^{\rm{U}}[n]D_{m}[n]}{  h_{m,k} [n] B_{m,k} }$,  and $x = \xi_{m,k}[n]$,
$\frac{{\partial}^2 \Tilde{\mathcal{W}}_{6}[n]}{{\partial  \xi^{2}_{m,k}[n]}} > 0$ can be obtained. Thus $\Tilde{\mathcal{W}}_{6}[n]$ is a convex function with respect to variables $\xi_{m,k}[n]$.
Similarly, $\frac{{\partial}^2 \Tilde{\mathcal{W}}_{6}[n]}{{\partial  \xi^{2}_{m,{\rm{LEO}}}[n]}} > 0$ can also be obtained, and $\Tilde{\mathcal{W}}_{6}[n]$ is a convex function with respect to variables $\xi_{m,{\rm{LEO}}}[n]$. This proof is completed.
\end{proof}

\end{appendices}

\bibliographystyle{IEEEtran}
\bibliography{myref}

@ARTICLE{Diao2020Fairness-Aware,
	author={Diao, Xianbang and Wang, Meng and Zheng, Jianchao and Cai, Yueming},
	journal={IEEE Access}, 
	title={Fairness-Aware Offloading and Trajectory Optimization for Multi-{UAV} Enabled Multi-Access Edge Computing}, 
	year={2020},
	volume={8},
	number={},
	pages={124359-124370}}

@ARTICLE{Shen2018Fractional1,
	author={Shen, Kaiming and Yu, Wei},
	journal={IEEE Trans. Signal Process.}, 
	title={Fractional Programming for Communication Systems—{Part} {I}: Power Control and Beamforming}, 
	year={May 2018},
	volume={66},
	number={10},
	pages={2616-2630}}

@ARTICLE{Zeng2023Joint,
  author={Zeng, Yaoping and Chen, Shisen and Cui, Yanpeng and Yang, Jie and Fu, Yinjuan},
  journal={IEEE Internet Things J.}, 
  title={Joint Resource Allocation and Trajectory Optimization in {UAV}-Enabled Wirelessly Powered {MEC} for Large Area}, 
  year={Sep. 2023},
  volume={10},
  number={17},
  pages={15705-15722}}

@ARTICLE{Zhou2021Deep,
  author={Zhou, Conghao and Wu, Wen and He, Hongli and Yang, Peng and Lyu, Feng and Cheng, Nan and Shen, Xuemin},
  journal={IEEE Trans. Wireless Commun.}, 
  title={Deep Reinforcement Learning for Delay-Oriented {IoT} Task Scheduling in {SAGIN}}, 
  year={Feb. 2021},
  volume={20},
  number={2},
  pages={911-925}}

@ARTICLE{Zhang2014Dynamic,
  author={Zhang, Ning and Liang, Hao and Cheng, Nan and Tang, Yujie and Mark, Jon W. and Shen, Xuemin Sherman},
  journal={IEEE J. Sel. Areas Commun.}, 
  title={Dynamic Spectrum Access in Multi-Channel Cognitive Radio Networks}, 
  year={Nov. 2014},
  volume={32},
  number={11},
  pages={2053-2064}}

@ARTICLE{Liu2023Energy-Efficient,
  author={Liu, Yi and Jiang, Li and Qi, Qi and Xie, Shengli},
  journal={IEEE Internet Things J.}, 
  title={Energy-Efficient Space-Air-Ground Integrated Edge Computing for Internet of Remote Things: A Federated {DRL} Approach}, 
  year={Mar. 2023},
  volume={10},
  number={6},
  pages={4845-4856}}

@INPROCEEDINGS{Zhang2023Learning,
  author={Zhang, Xuhui and Liu, Wenchao and Xing, Huijun and Jin, Zhenzhen and Zang, Weilin and Wang, Shuqiang and Shen, Yanyan and Xue, Liang},
  booktitle={Proc. IEEE Int. Conf. CYBER Technol. Autom., Control, Intell. Syst., CYBER}, 
  title={Learning to Hybrid Offload in Space-Air-Ground Integrated Mobile Edge Computing for {IoT} Networks}, 
  year={Qinhuangdao, China, Jul. 2023},
  volume={},
  number={},
  pages={836-841}}

@ARTICLE{Zeng2017Energy-Efficient,
  author={Zeng, Yong and Zhang, Rui},
  journal={IEEE Trans. Wireless Commun.}, 
  title={Energy-Efficient {UAV} Communication With Trajectory Optimization}, 
  year={Jun. 2017},
  volume={16},
  number={6},
  pages={3747-3760}}

@INPROCEEDINGS{Nguyen2022Joint,
  author={Nguyen, Minh Dat and Le, Long Bao and Girard, André},
  booktitle={Proc. IEEE Glob. Commun. Conf., GLOBECOM}, 
  title={Joint Computation Offloading, {UAV} Trajectory, User Scheduling, and Resource Allocation in {SAGIN}}, 
  year={Virtual, Brazil, Dec. 2022},
  volume={},
  number={},
  pages={5099-5104}}

@INPROCEEDINGS{Chen2022Energy,
  author={Chen, Bingchang and Li, Na and Li, Yan and Tao, Xiaofeng and Sun, Guen},
  booktitle={Proc. IEEE Wireless Commun. Netw. Conf., WCNC}, 
  title={Energy Efficient Hybrid Offloading in Space-Air-Ground Integrated Networks}, 
  year={Austin, TX, United states, Apr. 2022},
  volume={},
  number={},
  pages={1319-1324}}

@INPROCEEDINGS{2013MunozJoint,
  author={Muñoz, Olga and Pascual-Iserte, Antonio and Vidal, Josep},
  booktitle={Proc. Future Netw. Mob. Summit}, 
  title={Joint allocation of radio and computational resources in wireless application offloading}, 
  year={Lisbon, Portugal, Jul. 2013},
  volume={},
  number={},
  pages={1-10}}

@ARTICLE{2010LuoSemidefinite,
  author={Luo, Zhi-quan and Ma, Wing-kin and So, Anthony Man-cho and Ye, Yinyu and Zhang, Shuzhong},
  journal={IEEE Signal Process. Mag.}, 
  title={Semidefinite Relaxation of Quadratic Optimization Problems}, 
  year={May 2010},
  volume={27},
  number={3},
  pages={20-34}}

@ARTICLE{Liu2018Space-Air-Ground,
  author={Liu, Jiajia and Shi, Yongpeng and Fadlullah, Zubair Md. and Kato, Nei},
  journal={IEEE Commun. Surveys Tuts.}, 
  title={Space-Air-Ground Integrated Network: A Survey}, 
  year={4th Quart. 2018},
  volume={20},
  number={4},
  pages={2714-2741}}

@ARTICLE{Zhang2023Multiagent,
  author={Zhang, Senbai and Liu, Aijun and Han, Chen and Liang, Xiaohu and Xu, Xin and Wang, Guangyu},
  journal={IEEE Internet Things J.}, 
  title={Multiagent Reinforcement Learning-Based Orbital Edge Offloading in {SAGIN} Supporting Internet of Remote Things}, 
  year={Dec. 2023},
  volume={10},
  number={23},
  pages={20472-20483}}

@INPROCEEDINGS{Wang2021Survey,
  author={Wang, Heng and Xia, Xu and Song, Tangyijia and Xing, Yanxia},
  booktitle={Proc. IEEE/CIC Int. Conf. Commun. China, ICCC Workshops}, 
  title={Survey on Space-air-ground Integrated Networks in {6G}}, 
  year={Xiamen, China, Jul. 2021},
  volume={},
  number={},
  pages={315-320}}

@ARTICLE{Liu2020Task-Oriented,
  author={Liu, Jun and Du, Xinqi and Cui, Junhong and Pan, Miao and Wei, Debing},
  journal={IEEE Internet Things J.}, 
  title={Task-Oriented Intelligent Networking Architecture for the Space-Air-Ground-Aqua Integrated Network}, 
  year={Jun. 2020},
  volume={7},
  number={6},
  pages={5345-5358}}

@ARTICLE{Wang2024Hybrid,
  author={Wang, Xun and Chen, Hongbin and Tan, Fangqing},
  journal={IEEE Trans. Veh. Technol.}, 
  title={Hybrid {OMA/NOMA} Mode Selection and Resource Allocation in Space-Air-Ground Integrated Networks}, 
  year={Jan. 2025},
  volume={74},
  number={1},
  pages={699-713},
}

@INPROCEEDINGS{Dazhi2022Terminal-Aware,
  author={Dazhi, Michael N. and Al-Hraishawi, Hayder and Shankar, Bhavani and Chatzinotas, Symeon},
  booktitle={Proc. IEEE Glob. Commun. Conf., GLOBECOM Workshops}, 
  title={Terminal-Aware Multi-Connectivity Scheduler for Uplink Multi-Layer Non-Terrestrial Networks},
  year={Virtual, Brazil, Dec. 2022},
  volume={},
  number={},
  pages={1133-1139}}

@ARTICLE{Bakambekova2024On,
  author={Bakambekova, Adilya and Kouzayha, Nour and Al-Naffouri, Tareq},
  journal={IEEE Open J. Commun. Soc.}, 
  title={On the Interplay of Artificial Intelligence and Space-Air-Ground Integrated Networks: A Survey}, 
  year={Jul. 2024},
  volume={5},
  number={},
  pages={4613-4673}}

@ARTICLE{Wu2018Joint,
  author={Wu, Qingqing and Zeng, Yong and Zhang, Rui},
  journal={IEEE Trans. Wireless Commun.}, 
  title={Joint Trajectory and Communication Design for Multi-{UAV} Enabled Wireless Networks}, 
  year={Mar. 2018},
  volume={17},
  number={3},
  pages={2109-2121}}

@ARTICLE{Mao2017ASurvey,
  author={Mao, Yuyi and You, Changsheng and Zhang, Jun and Huang, Kaibin and Letaief, Khaled B.},
  journal={IEEE Commun. Surveys Tuts.}, 
  title={A Survey on Mobile Edge Computing: The Communication Perspective}, 
  year={4th Quart. 2017},
  volume={19},
  number={4},
  pages={2322-2358}}

@ARTICLE{Zhang2024A,
  author={Zhang, Siqi and Yi, Na and Ma, Yi},
  journal={IEEE Trans. Intell. Transp. Syst.}, 
  title={A Survey of Computation Offloading With Task Types}, 
  year={Aug. 2024},
  volume={25},
  number={8},
  pages={8313-8333}}

@ARTICLE{Mach2017Mobile,
  author={Mach, Pavel and Becvar, Zdenek},
  journal={IEEE Commun. Surveys Tuts.}, 
  title={Mobile Edge Computing: A Survey on Architecture and Computation Offloading}, 
  year={3rd Quart. 2017},
  volume={19},
  number={3},
  pages={1628-1656}}

@ARTICLE{Hu2023Joint,
  author={Hu, Zhenzhen and Zeng, Fanzi and Xiao, Zhu and Fu, Bin and Jiang, Hongbo and Xiong, Hailiang and Zhu, Yongdong and Alazab, Mamoun},
  journal={IEEE Trans. Veh. Technol.}, 
  title={Joint Resources Allocation and {3D} Trajectory Optimization for {UAV}-Enabled Space-Air-Ground Integrated Networks}, 
  year={Nov. 2023},
  volume={72},
  number={11},
  pages={14214-14229}}

@ARTICLE{9273074,
  author={Luo, Weiran and Shen, Yanyan and Yang, Bo and Wang, Shuqiang and Guan, Xinping},
  journal={IEEE Internet Things J.}, 
  title={Joint {3-D} Trajectory and Resource Optimization in Multi-{UAV}-Enabled {IoT} Networks With Wireless Power Transfer}, 
  year={May 2021},
  volume={8},
  number={10},
  pages={7833-7848},
}

@ARTICLE{10606316,
  author={Liu, Wenchao and Wang, Hao and Zhang, Xuhui and Xing, Huijun and Ren, Jinke and Shen, Yanyan and Cui, Shuguang},
  journal={IEEE Internet of Things Journal}, 
  title={Joint Trajectory Design and Resource Allocation in UAV-Enabled Heterogeneous MEC Systems}, 
  year={Oct. 2024},
  volume={11},
  number={19},
  pages={30817-30832}}

@ARTICLE{9953964,
  author={Chen, Qian and Meng, Weixiao and Han, Shuai and Li, Cheng and Chen, Hsiao Hwa},
  journal={IEEE Trans. on Cogn. Commun. Netw.}, 
  title={Effect of Intelligent Multi-Association in Civil Aircraft-Augmented {SAGIN}}, 
  year={Feb. 2023},
  volume={9},
  number={1},
  pages={223-238},
}

@ARTICLE{10458883,
  author={Mohamed, Ehab Mahmoud and Ahmed Alnakhli, Mohammad and Fouda, Mostafa M.},
  journal={IEEE Open J. Commun. Soc.}, 
  title={Joint {UAV} Trajectory Planning and {LEO}-Sat Selection in {SAGIN}}, 
  year={2024},
  volume={5},
  number={},
  pages={1624-1638},
}

@ARTICLE{9606690,
  author={Tang, Fengxiao and Hofner, Hans and Kato, Nei and Kaneko, Kazuma and Yamashita, Yasutaka and Hangai, Masatake},
  journal={IEEE J. Sel. Areas Commun.}, 
  title={A Deep Reinforcement Learning-Based Dynamic Traffic Offloading in Space-Air-Ground Integrated Networks ({SAGIN})}, 
  year={Jan. 2022},
  volume={40},
  number={1},
  pages={276-289},
}

@ARTICLE{9963692,
  author={Liu, Qiang and Fu, Meixia and Li, Wei and Xie, Jiagui and Kadoch, Michel},
  journal={IEEE Internet Things J.}, 
  title={{RIS}-Assisted Ambient Backscatter Communication for {SAGIN IoT}}, 
  year={Jun. 2023},
  volume={10},
  number={11},
  pages={9375-9384},
}

@ARTICLE{9918062,
  author={Tang, Fengxiao and Wen, Cong and Luo, Linfeng and Zhao, Ming and Kato, Nei},
  journal={IEEE J. Sel. Areas Commun.}, 
  title={Blockchain-Based Trusted Traffic Offloading in Space-Air-Ground Integrated Networks ({SAGIN}): A Federated Reinforcement Learning Approach}, 
  year={Dec. 2022},
  volume={40},
  number={12},
  pages={3501-3516},
}

@ARTICLE{10454605,
  author={Jia, Huaiqi and Wang, Ying and Wu, Wen},
  journal={IEEE Internet Things J.}, 
  title={Dynamic Resource Allocation for Remote {IoT} Data Collection in SAGIN}, 
  year={Jun. 2024},
  volume={11},
  number={11},
  pages={20575-20589},
}

@INPROCEEDINGS{10233456,
  author={Liu, Wenchao and Wang, Junyu and Xing, Huijun and Jin, Zhenzhen and Zhang, Xuhui and Shen, Yanyan},
  booktitle={Proc. IEEE/CIC Int. Conf. Commun. China, ICCC}, 
  title={Blockchain-Empowered Space-Air-Ground Integrated Networks for Remote Internet of Things}, 
  year={Dalian, China, Aug. 2023},
  volume={},
  number={},
  pages={1-6},
}

@ARTICLE{9964037,
  author={Meng, Xi and Zhang, Nan and Jian, Mengnan and Kadoch, Michel and Yang, Dacheng},
  journal={IEEE Internet Things J.}, 
  title={Channel Modeling and Estimation for Reconfigurable-Intelligent-Surface-Based {6G SAGIN IoT}}, 
  year={Jun. 2023},
  volume={10},
  number={11},
  pages={9273-9282},
}

@ARTICLE{8664595,
  author={Ren, Jinke and Yu, Guanding and He, Yinghui and Li, Geoffrey Ye},
  journal={IEEE Trans. Veh. Technol.}, 
  title={Collaborative Cloud and Edge Computing for Latency Minimization}, 
  year={May 2019},
  volume={68},
  number={5},
  pages={5031-5044},
}

@ARTICLE{8488502,
  author={Cao, Xiaowen and Wang, Feng and Xu, Jie and Zhang, Rui and Cui, Shuguang},
  journal={IEEE Internet Things J.}, 
  title={Joint Computation and Communication Cooperation for Energy-Efficient Mobile Edge Computing}, 
  year={Jun. 2019},
  volume={6},
  number={3},
  pages={4188-4200},
}

@ARTICLE{8960510,
  author={Wang, Feng and Xu, Jie and Cui, Shuguang},
  journal={IEEE Trans. Wireless Commun.}, 
  title={Optimal Energy Allocation and Task Offloading Policy for Wireless Powered Mobile Edge Computing Systems}, 
  year={Apr. 2020},
  volume={19},
  number={4},
  pages={2443-2459},
}

@INPROCEEDINGS{9417469,
  author={Zhang, Xuhui and Shen, Yanyan and Yang, Bo and Zang, Weilin and Wang, Shuqiang},
  booktitle={Proc. IEEE Wireless Commun. Netw. Conf., WCNC}, 
  title={{DRL} based Data Offloading for Intelligent Reflecting Surface Aided Mobile Edge Computing}, 
  year={Nanjing, China, Mar. 2021},
  volume={},
  number={},
  pages={1-7},
}

@ARTICLE{10373153,
  author={Liang, Bizheng and Fan, Rongfei and Hu, Han and Jiang, Hai and Xu, Jie and Zhang, Ning},
  journal={IEEE Trans. Veh. Technol.}, 
  title={Joint Task Offloading and Resource Allocation in Multi-User Mobile Edge Computing With Continuous Spectrum Sharing}, 
  year={May 2024},
  volume={73},
  number={5},
  pages={7234-7249},
}

@ARTICLE{10417719,
  author={Chen, Che and Gong, Shimin and Zhang, Wenjie and Zheng, Yifeng and Kiat, Yeo Chai},
  journal={IEEE Trans. on Cloud Comput.}, 
  title={DRL-Based Contract Incentive for Wireless-Powered and UAV-Assisted Backscattering MEC System}, 
  year={Jan.-Mar. 2024},
  volume={12},
  number={1},
  pages={264-276},
}

@ARTICLE{10440193,
  author={Huang, Chong and Chen, Gaojie and Xiao, Pei and Xiao, Yue and Han, Zhu and Chambers, Jonathon A.},
  journal={IEEE J. Sel. Areas Commun.}, 
  title={Joint Offloading and Resource Allocation for Hybrid Cloud and Edge Computing in {SAGINs}: A Decision Assisted Hybrid Action Space Deep Reinforcement Learning Approach}, 
  year={May 2024},
  volume={42},
  number={5},
  pages={1029-1043},
}

@ARTICLE{10342725,
  author={Nguyen, Minh Dat and Le, Long Bao and Girard, André},
  journal={IEEE Trans. on Cloud Comput.}, 
  title={Integrated Computation Offloading, {UAV} Trajectory Control, Edge-Cloud and Radio Resource Allocation in {SAGIN}}, 
  year={Jan.-Mar. 2024},
  volume={12},
  number={1},
  pages={100-115},
}

@ARTICLE{10579794,
  author={Du, Jianbo and Wang, Jiaxuan and Sun, Aijing and Qu, Junsuo and Zhang, Jianjun and Wu, Celimuge and Niyato, Dusit},
  journal={IEEE Internet Things J.}, 
  title={Joint Optimization in Blockchain- and {MEC}-Enabled Space–Air–Ground Integrated Networks}, 
  year={Oct. 2024},
  volume={11},
  number={19},
  pages={31862-31877},
}

@ARTICLE{9380358,
  author={Wang, Xianpeng and Yang, Laurence T. and Meng, Dandan and Dong, Mianxiong and Ota, Kaoru and Wang, Huafei},
  journal={IEEE Internet Things J.}, 
  title={Multi-{UAV} Cooperative Localization for Marine Targets Based on Weighted Subspace Fitting in {SAGIN} Environment}, 
  year={Apr. 2022},
  volume={9},
  number={8},
  pages={5708-5718},
}

@ARTICLE{8387798,
  author={Ren, Jinke and Yu, Guanding and Cai, Yunlong and He, Yinghui},
  journal={IEEE Trans. Wireless Commun.}, 
  title={Latency Optimization for Resource Allocation in Mobile-Edge Computation Offloading}, 
  year={Aug. 2018},
  volume={17},
  number={8},
  pages={5506-5519},
}

@ARTICLE{9508471,
  author={Hosseinian, Mohsen and Choi, Jihwan P. and Chang, Seok-Ho and Lee, Jungwon},
  journal={IEEE Aerosp. Electron. Syst. Mag.}, 
  title={Review of {5G NTN} Standards Development and Technical Challenges for Satellite Integration With the {5G} Network}, 
  year={Aug. 2021},
  volume={36},
  number={8},
  pages={22-31},
}

@ARTICLE{7901477,
  author={Tran, Tuyen X. and Hajisami, Abolfazl and Pandey, Parul and Pompili, Dario},
  journal={IEEE Commun. Mag.}, 
  title={Collaborative Mobile Edge Computing in {5G} Networks: New Paradigms, Scenarios, and Challenges}, 
  year={Apr. 2017},
  volume={55},
  number={4},
  pages={54-61},
}

@ARTICLE{10879508,
  author={Zhu, Wenwu and Deng, Xiaoheng and Gui, Jinsong and Zhang, Honggang and Min, Geyong},
  journal={IEEE Internet Things J.}, 
  title={Cost-Effective Task Offloading and Resource Scheduling for Mobile Edge Computing in {6G} Space-Air–Ground Integrated Network}, 
  year={Jun. 2025},
  volume={12},
  number={12},
  pages={19428-19442},
}

@ARTICLE{10891825,
  author={Li, Jialiuyuan and Shi, You and Dai, Chen and Yi, Changyan and Yang, Yuxiao and Zhai, Xiangping and Zhu, Kun},
  journal={IEEE Trans. Veh. Technol.}, 
  title={A Learning-Based Stochastic Game for Energy Efficient Optimization of {UAV} Trajectory and Task Offloading in Space/Aerial Edge Computing}, 
  year={Jun. 2025},
  volume={74},
  number={6},
  pages={9717-9733},
}

@ARTICLE{10947633,
  author={Xie, Wenxuan and Chen, Chen and Ju, Ying and Shen, Jun and Pei, Qingqi and Song, Houbing},
  journal={IEEE Trans. Intell. Transp. Syst.}, 
  title={Deep Reinforcement Learning-Based Computation Computational Offloading for Space–Air–Ground Integrated Vehicle Networks}, 
  year={May 2025},
  volume={26},
  number={5},
  pages={5804-5815},
}

@ARTICLE{10980172,
  author={Zhang, Xuhui and Xing, Huijun and Shen, Yanyan and Xu, Jie and Cui, Shuguang},
  journal={IEEE Trans. Wireless Commun.}, 
  title={Age of Information Minimization in {UAV}-Enabled {IoT} Networks via Federated Reinforcement Learning}, 
  year={Sep. 2025},
  volume={24},
  number={9},
  pages={7923-7939},
}

@ARTICLE{10972043,
  author={Zhang, Xuhui and Liu, Wenchao and Ren, Jinke and Xing, Huijun and Gui, Gui and Shen, Yanyan and Cui, Shuguang},
  journal={IEEE Internet Things J.}, 
  title={Latency Minimization for {UAV}-Enabled Federated Learning: Trajectory Design and Resource Allocation}, 
  year={Jul. 2025},
  volume={12},
  number={14},
  pages={27097-27112},
}

@ARTICLE{11134095,
  author={Zhang, Jianshan and Yang, Xu and Chen, Xing and Chen, Xiang and Yi, Xun and Khalil, Ibrahim and Niyato, Dusit},
  journal={IEEE Trans. Veh. Technol.}, 
  title={Energy-Efficient {UAV} Deployment and Computation Offloading in Space-Air-Ground Integrated Networks}, 
  year={to appear, 2025},
  volume={},
  number={},
  pages={},
}

@ARTICLE{11153428,
  author={Xu, Yue and Zheng, Linjiang and Wu, Xiao and Tang, Yi and Zhao, Min and Sun, Dihua},
  journal={IEEE Internet Things J.}, 
  title={Energy-Efficient Resource Allocation for Space-Air-Ground Integrated Vehicular Network}, 
  year={to appear, 2025},
  volume={},
  number={},
  pages={},
}

@ARTICLE{11145097,
  author={Hsu, Yi-Huai and Phan, Thi Thanh Tuyen},
  journal={IEEE Trans. Commun.}, 
  title={A {DRL}-Based Energy-Efficient Service Caching and Task Offloading Scheme for {6G} {MEC} {SAGINs}}, 
  year={to appear, 2025},
  volume={},
  number={},
  pages={},
}

@ARTICLE{11192086,
  author={Liu, Xiaomin and Peng, Yujie and Song, Xiaoqin and Song, Tiecheng},
  journal={IEEE Trans. Veh. Technol.}, 
  title={Latency-Aware Optimization of {UAV} Deployment, Computation Offloading, and Resource Allocation for {IoRT} in Space-Air-Ground Integrated Networks}, 
  year={to appear, 2025},
  volume={},
  number={},
  pages={},
}

@ARTICLE{11124243,
  author={Huynh, Dang Van and Khosravirad, Saeed R. and Cotton, Simon L. and Shin, Hyundong and Duong, Trung Q.},
  journal={IEEE Internet Things J.}, 
  title={Multi-Agent Reinforcement Learning for Optimal Resource Allocation in Space-Air-Ground Integrated Networks}, 
  year={to appear, 2025},
  volume={},
  number={},
  pages={},
}

@ARTICLE{11153721,
  author={Han, Zhe and Zheng, Guoqiang and Li, Chuanfeng and Shao, Hongxiang and Bai, Weiwei},
  journal={IEEE Sensors J.}, 
  title={{AoI}-Aware Scheduling Optimization for {WPT}-Enabled Space–Air–Ground Integrated Sensor Networks}, 
  year={2025},
  volume={25},
  number={20},
  pages={39082-39099},
}

@ARTICLE{8876867,
  author={Li, Gang and Cai, Jun},
  journal={IEEE Trans. Wireless Commun.}, 
  title={An Online Incentive Mechanism for Collaborative Task Offloading in Mobile Edge Computing}, 
  year={Jan. 2020},
  volume={19},
  number={1},
  pages={624-636},
}

@ARTICLE{8606230,
  author={Yi, Changyan and Cai, Jun and Su, Zhou},
  journal={IEEE Trans. Mobile Comput.}, 
  title={A Multi-User Mobile Computation Offloading and Transmission Scheduling Mechanism for Delay-Sensitive Applications}, 
  year={Jan. 2020},
  volume={19},
  number={1},
  pages={29-43},
}

@ARTICLE{10238695,
  author={Chen, Jiayuan and Yi, Changyan and Okegbile, Samuel D. and Cai, Jun and Shen, Xuemin},
  journal={IEEE Commun. Surveys Tuts.}, 
  title={Networking Architecture and Key Supporting Technologies for Human Digital Twin in Personalized Healthcare: A Comprehensive Survey}, 
  year={1st Quart. 2024},
  volume={26},
  number={1},
  pages={706-746},
}

@ARTICLE{10234396,
  author={Okegbile, Samuel D. and Cai, Jun and Zheng, Hao and Chen, Jiayuan and Yi, Changyan},
  journal={IEEE J. Sel. Areas Commun.}, 
  title={Differentially Private Federated Multi-Task Learning Framework for Enhancing Human-to-Virtual Connectivity in Human Digital Twin}, 
  year={Nov. 2023},
  volume={41},
  number={11},
  pages={3533-3547},
}

@ARTICLE{10540318,
  author={Yang, Yuye and Shi, You and Yi, Changyan and Cai, Jun and Kang, Jiawen and Niyato, Dusit and Shen, Xuemin},
  journal={IEEE Trans. Mobile Comput.}, 
  title={Dynamic Human Digital Twin Deployment at the Edge for Task Execution: A Two-Timescale Accuracy-Aware Online Optimization}, 
  year={Dec. 2024},
  volume={23},
  number={12},
  pages={12262-12279},
}
\end{document}